\documentclass[notitlepage, 12pt, showkeys]{revtex4-1}
\usepackage[T1]{fontenc}
\usepackage{amssymb}
\usepackage{graphicx}
\usepackage{color}
\usepackage{caption}
\usepackage{hyperref}
\usepackage{tikz}

\usepackage{amsmath}
\usepackage{braket}
\usepackage{graphicx}
\usepackage{appendix}
\usepackage{subcaption}

\begin{document}
\title{Finite time path field theory perturbative methods for local quantum spin chain quenches}

\author{Domagoj Kui\'{c}}

\email{Correspondence: dkuic@tvz.hr}

\affiliation{Zagreb University of Applied Sciences, Vrbik 8, 10000 Zagreb, Croatia}

\author{Alemka Knapp}

\email{aknapp@tvz.hr}

\affiliation{Zagreb University of Applied Sciences, Vrbik 8, 10000 Zagreb, Croatia}

\author{Diana \v{S}aponja-Milutinovi\'{c}}

\email{dsaponjam@tvz.hr}

\affiliation{Zagreb University of Applied Sciences, Vrbik 8, 10000 Zagreb, Croatia}

\date{September 5, 2024.}

\begin{abstract}
We discuss local magnetic field quenches using perturbative methods of finite time path field theory (FTPFT) in the following spin chains: Ising and XY in a transverse magnetic field. Their common characteristics are: (i) they are integrable via mapping to a second quantized noninteracting fermion problem; and (ii) when the ground state is nondegenerate (true for finite chains except in special cases), it can be represented as a vacuum of Bogoliubov fermions. By switching on a local magnetic field perturbation at finite time, the problem becomes nonintegrable and must be approached via numeric or perturbative methods. Using the formalism of FTPFT based on Wigner transforms (WTs) of projected functions, we show how to: (i) calculate the basic ``bubble'' diagram in the Loschmidt echo (LE) of a quenched chain to any order in the perturbation; and (ii) resum the generalized Schwinger--Dyson equation for the fermion two-point retarded functions in the ``bubble'' diagram, hence achieving the resummation of perturbative expansion of LE for a wide range of perturbation strengths under certain analyticity assumptions. Limitations of the assumptions and possible generalizations beyond it and also for other spin chains are further discussed.
\end{abstract}


\maketitle

\section{Introduction}
From the point of view of consideration of nonequilibrium dynamics of complex systems, studying them in different quench setups is the simplest way of bringing out and observing a variety of its different aspects. Quenches are realized by a sudden change of the Hamiltonian parameter(s); usually global ones, or in a setup like that applied in this work, a local change of transverse magnetic field (i.e., at a single site) of Ising and XY quantum spin chains. Both systems are integrable, and local perturbation of transverse magnetic field breaks the translational, but not $\mathbb{Z}_2$ parity symmetry ($\pi$ angle $z$-axis rotation symmetry) of these models. In this way, it renders the perturbed Hamiltonian to be nonintegrable. 

There is an abundance of works elucidating different aspects of global and local quenches in different spin chains \cite{Senegupta1, Silva1, Fagotti1, Rossini1, Rossini2, Campos1, Gambassi1, Guo1, Canovi1, Campos3, Calabrese1, Igloi1, Igloi2, Foini1, Riegler1, Polkovnikov1, Schuricht1, Calabrese2, Calabrese3, Fagotti2, Heyl1, Essler1, Mitra1, Marcuzzi1, Bertini1, Mitra2, Nadkishore1, Yang1, Zunkovic1, Jafari1, Paul1, Ding1, Porta1, Lupo1}. In quench setups in which the perturbed Hamiltonian remains integrable calculations are amenable to exact analytical treatment. Intimately depending on (non)integrability, relaxation and steady states of integrable spin chains show properties of nonthermalization, described in terms of Generalized Gibbs Ensembles~\cite{Essler1}, or long lived prethermal states in nonintegrable systems close to integrability~\cite{Mitra1, Marcuzzi1, Bertini1, Mitra2}, or in interacting systems with strong disorder, a failure to thermalize in any sense, exhibiting instead many body localization \cite{Nadkishore1, Yang1}. Furthermore, quenches reveal dynamical transitions not connected to equilibrium ones \cite{Zunkovic1, Jafari1}, dynamical behavior connected to a critical point~\cite{Senegupta1, Silva1, Gambassi1, Guo1, Foini1, Paul1}, and related topological properties \cite{Ding1, Porta1}. Local quenches, in combination with global ones, enable the study of the dynamical interplay between nonintegrability and integrability manifested in nonequilibrium response of the system~\cite{Lupo1}. Loschmidt echo (LE) as a measure of sensitivity of the closed system to an external perturbation, irreversibility and possible revivals of the initial state, is a natural observable for such purposes~\cite{Goussev1}. Studies implementing LE as a probe in global and local quenches have also proven it very useful in a context of the nonequilibrium work probability statistics and dynamical behavior near the critical point \cite{Silva1, Gambassi1} and the effect of frustration induced in antiferromagnetic (AFM) spin chains with odd number of sites through periodic boundary conditions (topological frustration) \cite{groupLE, Catalano1}. Having in mind that there are various proposals~\cite{Rossini3, Goold1, Knap1, Knap2, Dora1, Dorner1, Mazzola1} for measurements of LE renders such studies even more interesting.

Diagonalization of Ising and XY chain Hamiltonian is achieved through mapping (Jordan--Wigner transformation (JWT) followed by a Bogoliubov rotation) to second quantized noninteracting fermions \cite{Lieb1, Fabio-book}. This makes it possible to treat them, despite not being described by an appropriately constructed Lagrangian, as they are a free fermionic field on $1$d discretized spatial lattice. Local perturbation of the transverse magnetic field then introduces an external scattering potential that breaks the translational symmetry. This is reflected in the nonconservation of fermion number, momenta, and energy. We consider a quench scenario in which the system is initially in a nondegenerate ground state (formally represented as a vacuum of Bogoliubov fermions) and  the perturbation is switched on suddenly at $t=0$. Its subsequent time evolution up to finite time $t$ is determined by the nonintegrable perturbed Hamiltonian. Nonintegrability forces one to apply either exact numerical methods for calculating LE or approximate ones. In the context of topological frustration in Ising chain, results of LE calculations implementing very efficient and exact numerical approach based on \cite{Rossini3, Lieb1} have already been presented in \cite{groupLE}. Alternatively, one can try to implement approximate methods, for references see \cite{Goussev1}.

In this work, we show that natural setting for a resummation of the perturbative expansion of LE is found within the framework of finite time path field theory (FTPFT), where switching on the perturbation and a subsequent time evolution both occur at finite times. Such setting is inherently a nonequilibrium one, being different from standard applications of quantum field theory, where one assumes an adiabatic switching on starting at $t=-\infty$~\cite{Lancaster, Le Bellac, Coleman}. This approach has already been applied to finite time nonequilibrium processes, like production of photons in heavy ion collisions \cite{Dadic1}, or time dependent particle processes, like neutrino oscillations \cite{Dadic2} and kaon oscillations and decay. Issues of renormalization and causality (which are not present in this paper) have been also systematically addressed within the formalism \cite{Dadic3}. Our approach to the subject of this paper relies on the procedure \cite{Dadic4}, developed within the framework of FTPFT, for calculating convolution products of the projected two-point retarded Green's functions and their Wigner transforms (WTs). These functions appear in convolution products obtained through the application of Wick theorem to $n$-point function calculation at the $n$-th order of perturbative expansion of~LE.

The outline of the content of this paper is as follows. In Section \ref{The model}, we discuss a general model encompassing XY and Ising chains in transverse magnetic field with periodic boundary conditions (PBC). We introduce also a sudden local perturbation of the magnetic field. Description of diagonalization of the unperturbed Hamiltonian is redirected to Appendix \ref{diagonalization}, since it is rather well known from the literature \cite{Lieb1, Fabio-book}. Brief definition of LE and its interpretations are given in Section \ref{Loschmidt_echo}. Results of perturbative cumulant expansion of LE up to $n$-th order in the perturbation are presented in Section \ref{perturbative_calculations}. The structure of each order of perturbative cumulant expansion is described diagrammatically, in terms of ``direct'' and ``twisted'' ``bubble'' diagrams. We relate our calculation with the previous results on LE and work statistics in a local magnetic field quench of ferromagnetic (FM) Ising chain, evaluated in second-order cumulant expansion approximation in \cite{Silva1}. In Section~\ref{projected_functions_resummation}, we introduce basic FTPFT formalism of WTs of two-point retarded functions projected to finite time intervals, and of their convolution products. Then, we calculate convolution products of projected two-point retarded functions with inserted vertices appearing at the $n$-th order of perturbative expansion, i.e., ``bubble'' diagrams introduced in Section~\ref{perturbative_calculations}. Identifying each order of perturbative expansion as a sum of products of two terms of two generalized Schwinger--Dyson equations for two-point functions, we then resum the perturbative expansion of LE. In addition, we introduce two analyticity assumptions under which such resummation procedure is valid. Proof of the part of this argument is given in Appendix \ref{Poles_R_A}. Finally, in Section \ref{results} we present a comparison of the results of resummation of perturbative expansion with the exact numerical results obtained by diagonalization of the Hamiltonian and direct evaluation of the LE. Conclusions, possible generalizations to other spin chains, and prospects for future work are presented in Section \ref{conclusions}.

\section{The Model} \label{The model}

The Hamiltonian of the XY chain in a transverse magnetic field $h$ is 
\begin{equation}
H_0 =J \sum_{j=1}^N (\frac{1+\gamma}{2}\sigma_j^x \sigma_{j+1}^x +\frac{1-\gamma}{2}\sigma_j^y \sigma_{j+1}^y+ h\sigma_j^z) , \label{XY_Hamiltonian}
\end{equation}
where $\sigma_j^\alpha$, with $\alpha =x, y, z$, is a Pauli spin operator of the $j$-th spin in a chain. Parameter $J$ defines the energy scale, with $J \lessgtr 0$ corresponding to FM or AFM nearest neighbor couplings. We set it here to $J = + 1$. Setting the value of $xy$ plane anisotropy parameter $\gamma = \pm1$ reduces the model to a quantum Ising chain; $\gamma =0$ reduces the model to an XX chain (isotropic XY). Translational symmetry is imposed by PBC: $\sigma_{j+N}^\alpha \equiv \sigma_{j}^\alpha$, where $N$ is the number of spins in a chain. This is equivalent to a ring geometry. The model is integrable; it is first mapped by JWT to a spinless noninteracting fermionic model on a 1D lattice, then the second quantized fermionic Hamiltonian is diagonalized using a discrete Fourier transform, followed by a Bogoliubov transformation in momentum space (see \cite{Lieb1, Fabio-book} and Appendix~\ref{diagonalization} for details).

Next, we consider a sudden perturbation of the Hamiltonian $H_0 \rightarrow H_1 = H_0 + V$, such that a change $\delta h$ in the magnetic field strength $h$ is introduced instantaneously and only at one spin site, for example (and without losing generality because of translational symmetry of $H_0$) at $N$-th spin site, 
\begin{equation}
V = \delta h \sigma_N^z . \label{V_spins}
\end{equation}

Due 
 to local perturbation given by (\ref{V_spins}), translational symmetry of the model (\ref{XY_Hamiltonian}) is broken, but not $\mathbb{Z}_2$ parity symmetry, with the parity operator $\Pi^z = \bigotimes_{j=1}^N \sigma_j^z$ (see Appendix \ref{diagonalization}).

\section{Loschmidt Echo} \label{Loschmidt_echo}
For a quantum state $\ket{\psi }$ and a sudden quench of the Hamiltonian $H_0 \rightarrow H_1$ at $t=0$, LE is defined as $\mathcal{L}(t)= |\mathcal{G}(t)|^2$, where its complex amplitude is
\begin{equation}
\mathcal{G}(t) = \bra{{\psi }}e^{iH_0t}e^{-iH_1t}\ket{\psi } . \label{complex_amplitude_LE}
\end{equation}

Such a definition allows for following interpretations \cite{Goussev1} of $\mathcal{L}(t)$:
\begin{itemize}
\item[+] Overlap at time $t$ between two quantum states evolved from the same initial state $\ket{\psi }$ at $t=0$. One is evolved by the perturbed Hamiltonian $H_1$, and the other by the unperturbed Hamiltonian $H_0$. In this case, $\mathcal{L}(t)$ is a measure of sensitivity of time evolution of $\ket{\psi }$ to perturbations and is referred to also as fidelity. 
\item[+] Overlap between the initial state $\ket{\psi }$ and the state evolved in time, first by $H_1$ in the interval $(0,t)$, and then by $-H_0$ in the interval $(t,2t)$. For Hamiltonians $H_0$ and $H_1$ with time reversal symmetry; this second part of time evolution in $\mathcal{L}(t)$ is equivalent to time reversed backward evolution of the state $e^{-iH_1t}\ket{\psi }$ from time $t$ to $t=0$ by $H_0$. Hence, it is a measure of imperfect recovery of the initial state, i.e., irreversibility generated by the differences in forward and backward time evolution due to interactions with the environment and dephasing. 
\item[+] Time evolution operator from $t=0$ to time $t$ in the interaction picture is $U_I(t,0)= e^{iH_0t}e^{-iH_1t}$. In case of an eigenstate $\ket{\psi } = \ket{\psi_0}$ of the unperturbed Hamiltonian $H_0 \ket{\psi_0} = E_0\ket{\psi_0}$, LE $\mathcal{L}(t)$ is the overlap between the initial and time evolved state. In that context, it can be used as a measure of revival of the initial state.
\end{itemize}

\section{Perturbative Calculations} \label{perturbative_calculations}
All three interpretations of $\mathcal{L}(t)$ are equally valid, but the third one that utilizes the evolution operator in the interaction picture $U_I(t,0)$ is best fitted for time dependent perturbative calculations. By applying it, $\mathcal{G}(t) $ can be expressed as
\begin{eqnarray}
\mathcal{G}(t) & =& \bra{g_0}U_I(t,0)\ket{g_0} = \bra{g_0}Te^{-i\int_0^tV_I(t^\prime)dt^\prime}\ket{g_0} \nonumber \\
& = & \sum_{n=0}^\infty\frac{(-i)^n}{n!}\int_0^t dt_1 \dots \int_0^t dt_n \bra{g_0}T[V_I(t_1) \dots V_I(t_n)]\ket{g_0} , \label{G_perturbative}
\end{eqnarray}
where $T$ denotes the time ordered product of operators $V_I (t) = e^{iH_0t} V e^{-iH_0t}$. Here, we apply the approach introduced in \cite{Silva1}, and modify it for the XY model (\ref{XY_Hamiltonian}) with PBC (see also Appendix \ref{diagonalization}). We calculate $\mathcal{G}(t)$ for the ground state $H_0\ket{g_0}=E_0^{GS}\ket{g_0}$ of the unperturbed Hamiltonian. Using the cumulant expansion of (\ref{G_perturbative}), one obtains
\begin{eqnarray}
\mathcal{G}(t) = e^{\log[\mathcal{G}(t)]} = e^{ \sum_{n=1}^\infty\frac{(-i)^n}{n!}\int_0^t dt_1 \dots \int_0^t dt_n \bra{g_0}T[V_I(t_1) \dots V_I(t_n)]\ket{g_0}_C} \label{G_cumulant_exp}
\end{eqnarray}
where $C$ denotes only connected averages (or diagrams), in accordance with the linked-cluster theorem \cite{Coleman, Lancaster}. We assume that the ground state of the diagonalized second quantized Hamiltonian is nondegenerate. In particular, this is true for finite even or odd $N$ if $h^2 >|1-\gamma^2|$ (see Appendix \ref{diagonalization} and \cite{Fabio-book, Damski1} for further reference). The case of Ising chain corresponds to $\gamma = \pm 1$ and condition of nondegeneracy of the ground state in finite size chain reduces to $h \ne 0$.

Complex amplitude $\mathcal{G}(t)$ is usually written by the formula
\begin{equation}
\mathcal{G}(t) = e^{-i\delta E t}e^{ - f(t)} , \label{complex_Loschmidt_echo}
\end{equation}

To obtain the complex amplitude $\mathcal{G}(t)$ of LE we first express the local perturbation (\ref{V_spins}) in terms of the momentum space fermionic operators 
\begin{equation}
V = - \frac{\delta h}{N} \sum_{q, q^\prime \in \Gamma ^\pm } \left (c_q c_{q^\prime}^\dagger - c_q^\dagger c_{q^\prime} \right ) , \label{V_fermionic}
\end{equation}
where the minus sign in front of the sum depends only on the choice of JWT (\ref{JWT2}), and for JWT (\ref{JWT1}), it is replaced by a plus sign. Thus, when the choice of JWT (\ref{JWT1}) is necessary, all of the results from this point on are obtainable by a simple replacement of the sign of perturbation $\delta h \rightarrow -\delta h$.

Perturbation (\ref{V_fermionic}) generates the time evolution of ground states (\ref{gs_+_Bogoliubov}) and (\ref{gs_-_Bogoliubov}). By employing the periodicity $c_{q-N} = c_q$, we express it in terms of Bogoliubov operators
\begin{equation}
V = \frac{\delta h}{N} \sum_{q, q^\prime \in \Gamma ^\pm } \left( \begin{array}{cc} \eta _{q}^\dagger & \eta _{-q} \end{array} \right ) a_{q, q^\prime} \left( \begin{array}{c} \eta _{q^\prime} \\ \eta _{-q^\prime}^\dagger \end{array} \right ) . \label{V_Bogoliubov}
\end{equation}

Matrix $a_{q, q^\prime }$ depends on Bogoliubov angles,
\begin{equation}
a_{q, q^\prime} = \left( \begin{array}{c c} \cos (\theta _q + \theta _{q^\prime}) & -i\sin (\theta _q + \theta _{q^\prime})\\ i\sin (\theta _q + \theta _{q^\prime})& - \cos (\theta _q + \theta _{q^\prime}) \end{array} \right ) . \label{V_matrix}
\end{equation}

In terms of Bogoliubov fermions, perturbation is given by (\ref{V_Bogoliubov}) and (\ref{V_matrix}), so that fermions are created and/or annihilated in pairs. It is clear that the number, momenta $q$ and energy $\epsilon (q)$ of Bogoliubov fermions are not conserved. 

The calculation of the first-order term in the cluster expansion of $\log \mathcal{G}(t)$ is straightforward:
\begin{equation}
\bra{g_0^\pm } V \ket{g_0^\pm } = - \frac{\delta h}{N} \sum_{q \in \Gamma \pm } \cos 2\theta _q . \label{1th_ord_cluster}
\end{equation}

To calculate the second order of the expansion, time-ordered product $T[V_I(t_1)V_I(t_2)]$ is decomposed using Wick's theorem \cite{Coleman,Le Bellac, Lancaster}. Since the ground state $\ket{g_0^\pm }$ is a Bogoliubov vacuum, its expectation value of this time ordered product is simply a sum taken over all possible products of contractions of pairs of Bogoliubov operators $\eta _{q}(t) = e^{iH_0 t} \eta _{q}e^{-iH_0 t} $ and $\eta _{q}^\dagger (t) = e^{iH_0 t} \eta _{q}^\dagger e^{-iH_0 t} $ taken at different times. Equal time contractions are excluded as they correspond to disconnected diagrams, which do not contribute to cumulant expansion~(\ref{G_cumulant_exp}).

Up to the second-order $(\delta h)^2$ of the perturbation (\ref{V_spins}), $\delta E$ is equal to
\begin{equation}
\delta E_2 = - \frac{\delta h}{N}\sum_{q \in \Gamma ^{\pm }} \cos (2 \theta _q) - \frac{(\delta h)^2}{N^2}\sum_{q_1, q_2 \in \Gamma ^{\pm }}\frac{\sin^2 (\theta_{q_1} + \theta_{q_2}) }{\Lambda _{q_1} + \Lambda _{q_2}}, \label{deltaE}
\end{equation}
and $f(t)$ is given by 
\begin{equation}
f_2(t) = \frac{(\delta h)^2}{N^2}\sum_{q_1, q_2 \in \Gamma ^{\pm }} \frac{\sin^2 (\theta_{q_1} + \theta_{q_2}) }{2(\Lambda _{q_1} + \Lambda _{q_2})^2} \left ( 1 - e^{-i2(\Lambda _{q_1} + \Lambda _{q_2})t} \right ) . \label{fun_t}
\end{equation}

For $N$ even and $h \ne 0$ and, also for $N$ odd and $h < 0$, sums in Formulas (\ref{deltaE}) and (\ref{fun_t}) are over the set of fermion momenta $\Gamma^+ = \{q = 2\pi(k + \frac{1}{2})/N : k =0, 1, \dots, N - 1 \} $. For $N$ odd and $h > 0$, the sums are over $\Gamma^- = \{q = 2\pi k /N : k =0, 1, \dots, N - 1 \}$. With the appropriate JWT definition of fermions, and then by applying a Bogoliubov transformation, ground state $\ket{g_0^{\pm}}$ is expressed as a Bogoliubov vacuum (see Appendix \ref{diagonalization}). Here, $+$ and $-$ denote the parity of the ground state. Its energy is 
\begin{equation}
E_0^{GS \pm } = -\sum_{q \in \Gamma^\pm }\epsilon (q)/2 . \label{en_gs}
\end{equation}

Excitation energies of Bogoliubov fermions are 
\begin{eqnarray}
\epsilon (q) = 2\Lambda _q = 2 \left [(h - \cos q)^2 +\gamma^2 \sin^2 q \right ]^{1/2}, & \quad \forall q \in \Gamma ^{\pm } \backslash \{0, \pi \} . \label{en_eps}
\end{eqnarray}

Exceptions are
\begin{equation}
\epsilon (q) = 2\Lambda _q = \left \{ \begin{array}{l@{\,,\qquad }l} 2(h - \cos q) & N \ \mathrm{even}, \{q = 0, \pi\} \in \Gamma ^{- } \\ - 2(h - \cos q) & N \ \mathrm{odd}, h < 0, \{q = 0\} \in \Gamma ^{- }, \{q=\pi\} \in \Gamma ^{+ } 
\\ 2(h - \cos q) & N \ \mathrm{odd}, h > 0, \{q = 0\} \in \Gamma ^{- }, \{q=\pi\} \in \Gamma ^{+ } \end{array} \right . . \label{excitation_energies}
\end{equation}

For completeness, we also point out that our choice of JWT, for $N$ even, or for $N$ odd and $h \gtrless 0$, is intended, so that the ground state is always represented as a Bogoliubov vacuum (see Appendix \ref{diagonalization}). Because of this, angle $\theta _q$ of Bogoliubov transformation that diagonalizes the Hamiltonian depends on the choice of JWT, and it is given by
\begin{equation}
e^{i2\theta _q} = \left \{ \begin{array}{l@{\,,\qquad }l} \frac{ h - \cos q - i\gamma \sin q}{\Lambda _q} & N \ \mathrm{even} \\ \frac{ - h + \cos q + i\gamma \sin q}{\Lambda _q} & N \ \mathrm{odd}, h < 0 
\\ \frac{h - \cos q - i\gamma \sin q}{\Lambda _q} & N \ \mathrm{odd}, h > 0 \end{array} \right . . \label{Bogoliubov_angle}
\end{equation}

Taking the thermodynamic limit $N \gg 1 $ in Formulas (\ref{deltaE}) and (\ref{fun_t}), after PBC was assumed initially in the diagonalization of Hamiltonian (\ref{XY_Hamiltonian}), is qualitatively different procedure than the one applied by Silva \cite{Silva1}, in second-order cumulant expansion of LE for a FM version of the Ising model in a transverse magnetic field near a critical point. Procedure taken in \cite{Silva1} implies that, for a large system, effects due to the existence of a boundary of the system are negligible on the rest of chain. In that way, for a large system, the model is integrable and there is no need to introduce PBC. On the other hand, we have introduced PBC here in order to achieve integrability for finite chains. Of course, even without PBC, momenta $q$ and $q \pm 2\pi$ are physically identical, because, in a discrete chain, physically distinguishable momenta are restricted to the first Brillouin zone and, thus, overcounting of identical modes is avoided. But the difference how boundary conditions are treated by the two procedures has a qualitative effect and, according to \cite{Silva1}, 
 the excitation energies of Bogoliubov fermions are always nonnegative, while in the procedure used here they can be negative for $q=0, \pi$ modes (see (\ref{en_gs}), (\ref{en_eps}) and (\ref{excitation_energies}), and also Appendix \ref{diagonalization} for the details). 

For $N$ finite and odd, empty negative energy fermionic modes, $q=0$, for $0 < h < 1$, and $q = \pi$, for $-1 < h < 0$, appear in Bogoliubov vacuum representation of nondegenerate ground state of AFM phase of Ising and XY chains with PBC. They are responsible for algebraical closing of the gap ($\propto 1/N^2 $) between the nondegenerate ground state and a band of states on top of it in thermodynamic limit. The closing of the gap manifests itself in rich phenomenology of topological frustration discussed in AFM chains with PBC and odd number of spins \cite{Dong1, Dong2, group1, group2, group3, group4, group5, group6, group7, Odavic1}.

On the other hand, if we were to replace $\epsilon (q) $ given by \ref{en_eps} and \ref{excitation_energies} with nonnegative $\epsilon (q)= 2\left [(h - \cos q)^2 + \gamma^2 \sin^2 q \right ]^{1/2}$, and then take the thermodynamic limit in formula~(\ref{deltaE}) and (\ref{fun_t}), one would straightforwardly obtain what would be a result of the procedure used in \cite{Silva1}, if it were applied on the AFM version of the models discussed above. However, proceeding in such a way, intrinsic and nontrivial PBC induced effects would be completely lost. This would make our calculations of LE insensitive to qualitatively distinctive important features in the time-dependent behavior of AFM chains due to topological frustration induced by PBC in chains with odd number of spins \cite{groupLE}. 

In the next step, continuing beyond calculations in \cite{Silva1}, we obtain $n$-th order perturbative expansion of (\ref{G_cumulant_exp}), written as $\mathcal{G}_n = \mathcal{G}_{n-1} e^{-i\delta E_n t}e^{ - f_n(t)} $. $ \mathcal{G}_{{\mathrm n - 1}}$ is the $(n-1)$ order expansion. Methods used in calculations are explained in detail in the rest of this section and in Section \ref{projected_functions_resummation}. We obtain (rather bulky looking) expressions\vspace{-9pt}
\begin{eqnarray}
\delta E_{\mathrm{n}} & = & - \frac{2^{n-1}(\delta h)^n}{N^n}\sum_{q_{n+1}, q_2 , \dots, q_n \in \Gamma ^{\pm }}\sum_{j=2}^{\mathrm{n}} \sum_{k=2}^{j}\sum_{l=j+1}^{\mathrm{n+1}} \nonumber\\
&& \left [\sin (\theta_{q_{n+1}} + \theta_{q_2}) \frac{\cos (\theta_{q_2} + \theta_{q_3})\cos (\theta_{q_3} + \theta_{q_4}) \cdots \cos (\theta_{q_{j-1}} + \theta_{q_{j}}) }{2(\Lambda _{q_k} - \Lambda _{q_2})2(\Lambda _{q_k} - \Lambda _{q_3}) \cdots 2(\Lambda _{q_k} - \Lambda _{q_{j-1}})2(\Lambda _{q_k} - \Lambda _{q_j})} \sin (\theta_{q_{j}} + \theta_{q_{j+1}}) \right . \nonumber\\
&&\left. \times \frac{ \cos (\theta_{q_{j+1}} + \theta_{q_{j+2}}) \cos (\theta_{q_{j+2}} + \theta_{q_{j+3}}) \cdots \cos (\theta_{q_n} + \theta_{q_{n+1}})}{2(\Lambda _{q_l} - \Lambda _{q_{j+1}})2(\Lambda _{q_l} - \Lambda _{q_{j+2}}) \cdots 2(\Lambda _{q_l} - \Lambda _{q_{n}})2(\Lambda _{q_{l}} - \Lambda _{q_{n+1}})}\right ] \frac{1}{2(\Lambda _{q_{k}} + \Lambda _{q_l})} , \label{nth_ord_cum2} 
\end{eqnarray}
and\vspace{-9pt}
\begin{eqnarray}
f_{\mathrm{n}} & = & \frac{2^{n-1}(\delta h)^n}{N^n}\sum_{q_{n+1}, q_2 , \dots, q_n \in \Gamma ^{\pm }}\sum_{j=2}^{\mathrm{n}} \sum_{k=2}^{j}\sum_{l=j+1}^{\mathrm{n+1}} \nonumber\\
&& \left [\sin (\theta_{q_{n + 1}} + \theta_{q_2}) \frac{\cos (\theta_{q_2} + \theta_{q_3})\cos (\theta_{q_3} + \theta_{q_4}) \cdots \cos (\theta_{q_{j-1}} + \theta_{q_{j}}) }{2(\Lambda _{q_k} - \Lambda _{q_2})2(\Lambda _{q_k} - \Lambda _{q_3}) \cdots 2(\Lambda _{q_k} - \Lambda _{q_{j-1}})2(\Lambda _{q_k} - \Lambda _{q_j})} \sin (\theta_{q_{j}} + \theta_{q_{j+1}}) \right . \nonumber\\
&&\left. \times \frac{ \cos (\theta_{q_{j+1}} + \theta_{q_{j+2}}) \cos (\theta_{q_{j+2}} + \theta_{q_{j+3}}) \cdots \cos (\theta_{q_n} + \theta_{q_{n+1}})}{2(\Lambda _{q_l} - \Lambda _{q_{j+1}})2(\Lambda _{q_l} - \Lambda _{q_{j+2}}) \cdots 2(\Lambda _{q_l} - \Lambda _{q_{n}})2(\Lambda _{q_{l}} - \Lambda _{q_{n+1}})}
\right ] \frac{1-e^{-i 2(\Lambda _{q_k} + \Lambda _{q_{l}})t}}{(2(\Lambda _{q_k} + \Lambda _{q_{l}}))^2} . \label{nth_ord_cum4}
\end{eqnarray}

In both of these expressions, divergent factors $2(\Lambda _{q_k} - \Lambda _{q_k})$ and $2(\Lambda _{q_l} - \Lambda _{q_l})$, with the same summation indices in denominators, are absent as factors from the outset. Instead of them, there is factor $1$ in the denominator. Also, in a term with index $j=n$, products in this term end with $\sin (\theta_{q_{n}} + \theta_{q_{n+1}}) $, and there are no further factors inside the square brackets. This can be checked in the results from Section \ref{projected_functions_resummation}.

We can better summarize and explain perturbative results by utilizing a diagrammatic picture. An $n$-th order term of perturbative expansion (\ref{G_cumulant_exp}), as it is written here, is equal to 
\begin{equation}
\frac{(-i)^n}{n!}\int_0^t dt_1 \dots \int_0^t dt_n \bra{g_0^{\pm}}T[V_I(t_1) \dots V_I(t_n)]\ket{g_0^{\pm}}_C = -i \delta E_\mathrm{n} t -f_{\mathrm{n}}(t) . \label{nth_ord_cum}
\end{equation}

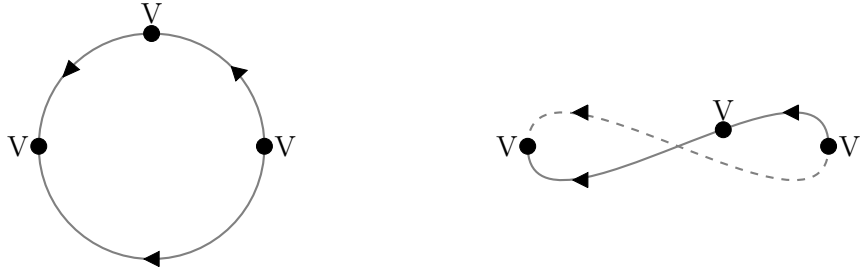
\begin{figure}
\begin{tikzpicture}
\draw[gray, thick] (0,0) circle (1.5);
\filldraw[black] (-1.5,0) circle (3pt) node[anchor=east] {V};
\filldraw[black] (1.5,0) circle (3pt) node[anchor=west] {V};
\filldraw[black] (0,1.5) circle (3pt) node[anchor=south] {V};

\draw [black, fill=black] (0.1,-1.6)--(-0.1,-1.5)--(0.1,-1.4)--(0.1,-1.6);
\draw [black, fill=black] (1.06,1.06)--(1.12,0.87)--(1.27,1.01)--(1.06,1.06);
\draw [black, fill=black] (-1.18, 0.92)--(-0.96,1.02)--(-1.1,1.14)--(-1.18,0.92);

\draw[gray, thick, dashed] (5,0) to [out=90,in=-90] (9,0); 
\draw[gray, thick] (5,0) to [out=-90,in=90] (9,0);
\filldraw[black](5,0) circle (3pt) node[anchor=east] {V};
\filldraw[black](9,0) circle (3pt) node[anchor=west] {V};

\filldraw[black] (7.6,0.22) circle (3pt) node[anchor=south] {V};

\draw [black, fill=black] (5.8,0.55)--(5.6,0.45)--(5.8,0.35)--(5.8,0.55);
\draw [black, fill=black] (5.8,-0.55)--(5.6,-0.45)--(5.8,-0.35)--(5.8,-0.55);
\draw [black, fill=black] (8.6,0.55)--(8.4,0.45)--(8.6,0.35)--(8.6,0.55);
\end{tikzpicture}
\caption{``Direct'' 
 and ``twisted'' ``bubble'' diagrams (left and right part of the figure, respectively) 
appearing at the $3$rd order of perturbative expansion of (\ref{G_cumulant_exp}). A dashed line is used to indicate that there is no intersection with a solid line. 
  V is an element of perturbation matrix (\ref{V_Bogoliubov}) and (\ref{V_matrix}) and, hence, a vertex in two strings of vacuum contractions of time ordered product of fermionic operators comprising a ``bubble''~diagram.}\label{fig1}
\end{figure}

It consists of two types of ``bubble'' diagrams appearing at second and all orders higher than that. These are ``direct'' and ``twisted'' ``bubble'' diagrams. At $n$-th order, each one has, in addition to two vertices appearing at the second order, $n-2$ additional vertex insertions. Vertex is an element of perturbation matrix (\ref{V_Bogoliubov}) and (\ref{V_matrix}). For example, a third order term in perturbative expansion (\ref{G_cumulant_exp}) consists of ``direct'' and ``twisted'' diagrams with one additional vertex insertion as illustrated on Figure \ref{fig1}. 

A ``twist'' means an inversion of Bogoliubov fermionic operators in the vertex with least time, joining the two strings of vacuum (i.e., ground state) contractions of time ordered product of operators in a ``bubble'' diagram. The least time vertex is always that element of the perturbation matrix, given by (\ref{V_Bogoliubov}) and (\ref{V_matrix}), that contains only creation operators; it is the element proportional to $ -i (\delta h /N) \sin (\theta _q + \theta _{q^\prime})$. The vertex with the greatest time, at the other joint of two strings of vacuum contractions, always contains only annihilation operators, and it is the element of perturbation matrix (\ref{V_Bogoliubov}) and (\ref{V_matrix}) proportional to $ i (\delta h /N ) \sin (\theta _q + \theta _{q^\prime})$. All vertices have a pair of equal time fermionic operators, so ``twisting'' a ``bubble'' diagram and making vacuum contractions in a time ordered product, due to Wick theorem, does not change the sign in front; it is the same as for a ``direct'' ``bubble'' diagram. 

All other intermediate vertices in two joined strings of a ``direct'' or a ``twisted'' ``bubble'' diagram are elements of (\ref{V_Bogoliubov}) and (\ref{V_matrix}) that contain a combination of annihilation and creation operator, i.e., the elements proportional to $(\delta h /N) \cos (\theta _q + \theta _{q^\prime})$ or $-(\delta h /N) \cos (\theta _q + \theta _{q^\prime})$. Again, time ordering and contracting is responsible that all such vertices obtain a $+$ sign in front of $(\delta h /N ) \cos (\theta _q + \theta _{q^\prime})$ vertex factor in a diagram. Equal time contractions are excluded, since they generate disconnected diagrams and cumulant expansion (\ref{G_cumulant_exp}) contains only connected diagrams. Finally, due to translational and other symmetries of the Hamiltonian (\ref{XY_Hamiltonian}), we have that $\Lambda _q = \Lambda_{-q} = \Lambda_{N-q}$ and $\theta _q = - \theta_{-q} = - \theta_{N-q}$. This is obvious from (\ref{en_eps}), (\ref{excitation_energies}), and (\ref{Bogoliubov_angle}). 

As a consequence of all the above facts, the total of $2^{n-1} n! \sum_{k=0} ^{n-2} {n -2\choose n -2 - k}$ ``bubble'' diagrams appearing at the $n$-th order of perturbative expansion (\ref{nth_ord_cum}), $2^{n-2} n! \sum_{k=0} ^{n-2} {n -2\choose k}$ ``direct'' ones and $2^{n-2} n! \sum_{k=0} ^{n-2} {n -2\choose k}$ ``twisted'' ones, have the form given in summary by the sums (\ref{nth_ord_cum2}) and (\ref{nth_ord_cum4}). Each term in these sums corresponds to $2^{n-1} n! $  
properly time ordered ``direct'' or ``twisted'' ``bubble'' diagrams.

\section{Projected Functions and Resummation} \label{projected_functions_resummation}

Vacuum contractions of Bogoliubov fermionic annihilation $\eta _{q}(t) = e^{iH_0 t} \eta _{q}e^{-iH_0 t} $ and creation operators $\eta _{q}^\dagger (t) = e^{iH_0 t} \eta _{q}^\dagger e^{-iH_0 t} $ appearing in (\ref{nth_ord_cum}) are essentially retarded two-point Green's functions,
\begin{equation}
G_{R, q} (t_1-t_2)= \bra{g_0^{\pm}}\eta _{q}(t_1) \eta _{q}^\dagger(t_2) \ket{g_0^{\pm}}\theta(t_1 - t_2) = e^{-i 2\Lambda _{q} (t_1-t_2)}\theta(t_1 - t_2) . \label{Green_2point_retarded}
\end{equation}

The WT of two-point function (\ref{Green_2point_retarded}), i.e., the Fourier transform with respect to $s_0 = t_1 - t_2$, has a pole in the lower complex semiplane,
\begin{equation}
G_{R,q}(p_0) = \frac{i}{p_0 - 2\Lambda _{q} + i\varepsilon} . \label{WT_Green_2point_retarded}
\end{equation}

The time-ordered product in (\ref{nth_ord_cum}) of perturbations (\ref{V_Bogoliubov}) is a $n$-point function. According to the Wick theorem, it consists of products of retarded two-point functions (\ref{Green_2point_retarded}). By applying the Wick theorem to obtain the time-ordered product and then taking $n$ time integrals in (\ref{nth_ord_cum}), we obtain convolution products of retarded two point functions (\ref{Green_2point_retarded}), with inserted vertices~(\ref{V_Bogoliubov}) and (\ref{V_matrix}). To be more illustrative, the time integrals in (\ref{nth_ord_cum}) make two strings of convolution products, which join at the vertex with greatest time and at the vertex with the lowest time of the two strings, making a ``bubble'' diagram.

Different from calculations of diagrams and $n$-point functions in a field theory setup with interactions switched on adiabatically from $t= - \infty$, in the quench-like scenario described in Section \ref{Loschmidt_echo}, perturbation is switched on suddenly at $t=0$. Time evolution of a perturbed ground state is then followed up until a finite time $t$. Hence, all time integrals in convolutions of retarded two-point functions in (\ref{nth_ord_cum}) are over finite interval with times in $0 \leq t_1, t_2, \dots, t_n \leq t$. 

These convolutions in reality are convolutions of retarded two-point functions (\ref{Green_2point_retarded}) that are projected to a finite time interval of evolution $0 \leq t_1, t_2 \leq t$,
\begin{equation}
G_{t, R, q} (t_1-t_2) = \theta(t) \theta(t - t_1) \theta(t_1) \theta(t - t_2)\theta(t_2) e^{-i 2\Lambda _{q} (t_1-t_2)}\theta(t_1 - t_2) . \label{Green_2point_retarded_projected}
\end{equation}

Product of $\theta$-functions $\theta(t) \theta(t - t_1) \theta(t_1) \theta(t - t_2)\theta(t_2)$ in front of $G_{R, q} (t_1-t_2)$ is the projector. Taking its WT, one obtains the important property
\begin{equation}
G_{t, R, q} (p_o) = \int_{-\infty}^{\infty} dp_0^\prime \frac{\theta(t)}{\pi}\frac{\sin(2t(p_0 - p_0^\prime))}{p_0 - p_0^\prime}G_{R,q}(p_o^\prime) , \label{WT_Green_2point_retarded_projected} 
\end{equation}
where $G_R(p_o)$ is the WT of (\ref{Green_2point_retarded}) given by (\ref{WT_Green_2point_retarded}). The factor in front of it is WT of the projector. This is also true for WTs of other projected functions. The theory of projected functions, their WTs, and their convolution products, developed within the framework of FTPFT, is described in detail in reference \cite{Dadic4}. It is sufficient for the purpose of this paper to draw attention to two important properties:

\begin{itemize}
\item[+] WT of a projected function (\ref{Green_2point_retarded_projected}) is 
\begin{equation}
G_{t, R, q} (p_o) = \frac{i(1 - e^{i2t(p_0 - 2\Lambda _{q} + i\varepsilon)})}{p_0 - 2\Lambda _{q} + i\varepsilon } .
\end{equation}

It is evident that $G_{t, R, q} (t_1-t_2)$ is retarded, like $G_{R, q} (t_1-t_2)$ .

\item[+] Convolution product of $n$ projected retarded functions (\ref{Green_2point_retarded_projected})
\begin{equation}
C_t (t_1 - t_n) = \int_{-\infty}^{\infty} dt_2 \dots \int_{-\infty}^{\infty} dt_{n-1}G_{t, R, q_1} (t_1-t_2) \cdots G_{t, R, q_n} (t_{n-1}-t_n) ,
\end{equation}
has a WT that obeys the rule
\begin{eqnarray}
C_t (p_0) = \int_{-\infty}^{\infty} dp_0^\prime \frac{\theta(t)}{\pi}\frac{\sin(2t(p_0 - p_0^\prime))}{p_0 - p_0^\prime}G_{R,q_1}(p_o^\prime) \cdots G_{R,q_2}(p_o^\prime) . \label{WT_convolution_product}
\end{eqnarray}

Here, $G_{R,q_1}(p_o^\prime) \cdots G_{R,q_2}(p_o^\prime)$ is a WT of a convolution product of $n$ retarded functions (\ref{Green_2point_retarded}). So, the same WT rule is obeyed by convolution products of projected retarded functions (\ref{Green_2point_retarded_projected}), as for the functions themselves. 
\end{itemize}

We now use the rule (\ref{WT_convolution_product}) to calculate two convolution products of projected retarded two point functions $G_{t, R, q} (t_1-t_2)$ in (\ref{nth_ord_cum}), joining at greatest and lowest time vertices $\pm i (\delta h /N)\sin (\theta _q + \theta _{q^\prime})$. The intermediate time vertices $ (\delta h /N)\cos (\theta _q + \theta _{q^\prime})$ are inserted between retarded two-point functions. We obtain\vspace{-3pt}
\begin{eqnarray}
&& \frac{(-i)^n}{n!} \int_0^t dt_1 \dots \int_0^t dt_n \bra{g_0^{\pm}}T[V_I(t_1) \dots V_I(t_n)]\ket{g_0^{\pm}}_C \cr \nonumber \\
&& = \frac{(-i)^n}{n!} \int_{0}^t dt_1 \int_{0}^t dt_n \theta(t_1 - t_n) \left (- \frac{2^{n-1}n!(\delta h)^n i^n}{N^n}\right) \sum_{j=2}^{\mathrm{n}} \sum_{q_1, q_2 , \dots, q_n \in \Gamma ^{\pm }} \cr \nonumber\\ 
&& \left \{\sin (\theta_{q_1} + \theta_{q_2}) \left [ \frac{1}{2\pi}\int_{-\infty}^{\infty} dp_0 e^{-ip_o(t_1 - t_n)} \int_{-\infty}^{\infty} dp_0^\prime \frac{\theta(t)}{\pi}\frac{\sin(2t(p_0 - p_0^\prime))}{p_0 - p_0^\prime}\right . \right . \cr \nonumber\\
&& \left. \left . \frac{\cos (\theta_{q_2} + \theta_{q_3})\cos (\theta_{q_3} + \theta_{q_4}) \cdots \cos (\theta_{q_{j-1}} + \theta_{q_{j}})}{(p_0^\prime - 2\Lambda _{q_2} + i\varepsilon) (p_0^\prime - 2\Lambda _{q_3} + i\varepsilon) \cdots (p_0^\prime - 2\Lambda _{q_{j-1}} + i\varepsilon)(p_0^\prime -2 \Lambda _{q_j} + i\varepsilon)} \right ] \right . \cr \nonumber\\
&& \left . \sin (\theta_{q_{j}} + \theta_{q_{j+1}}) \left [ \frac{1}{2\pi}\int_{-\infty}^{\infty} dk_0 e^{-ik_o(t_1 - t_n)} \int_{-\infty}^{\infty} dk_0^\prime \frac{\theta(t)}{\pi}\frac{\sin(2t(k_0 - k_0^\prime))}{k_0 - k_0^\prime} \right . \right . \cr \nonumber\\
&&\left. \left. \frac{ \cos (\theta_{q_{1}} + \theta_{q_{n}}) \cos (\theta_{q_{n}} + \theta_{q_{n-1}}) \cdots \cos (\theta_{q_{j+2}} + \theta_{q_{j+1}})}{(k_0^\prime - 2\Lambda _{q_1} + i\varepsilon)(k_0^\prime - 2\Lambda _{q_n} + i\varepsilon) \cdots (k_0^\prime - 2\Lambda _{q_{j+2}} + i\varepsilon)(k_0^\prime -2 \Lambda _{q_{j+1}} + i\varepsilon)} \right ] \right \} . 
 \label{n_ord_cumulant_convolution}
\end{eqnarray}

In 
 the term of a sum with index $j=n$, in place of $\sin (\theta_{q_{n}} + \theta_{q_{n+1}}) $, there is $\sin (\theta_{q_{n}} + \theta_{q_{1}}) $ factor. Also, under the integrals inside the second square brackets, there is only $1/(k_0^\prime - 2\Lambda _{q_1} + i\varepsilon)$ 
behind the WT of the projector. Similarly, in the $j=2$ term, under the integrals inside the first square brackets, there is only $1/(p_0^\prime - 2\Lambda _{q_2} + i\varepsilon)$ behind the WT of the projector. Factor $2^{n-1} n!$ 
 in front of the sum is the number of ``bubble'' diagrams corresponding to the same term of the sum. Factor $i^n$ comes from factor $i$ in the WTs (\ref{WT_Green_2point_retarded}) of retarded two-point functions. 

In the next step, by taking the integral over $p_0$ or over $p_0^\prime$ in the first square bracket, and the integral over $k_0$ or over $k_0^\prime$ 
 in the second square bracket of (\ref{n_ord_cumulant_convolution}), one finally obtains
\begin{eqnarray}
&& \frac{(-i)^n}{n!} \int_0^t dt_1 \dots \int_0^t dt_n \bra{g_0^{\pm}}T[V_I(t_1) \dots V_I(t_n)]\ket{g_0^{\pm}}_C \cr \nonumber \\
&& = - \frac{2^{n-1}(\delta h)^n }{N^n} \int_{0}^t dt_1 \int_{0}^t dt_n \theta(t_1 - t_n) \sum_{j=2}^{\mathrm{n}} \sum_{q_1, q_2 , \dots, q_n \in \Gamma ^{\pm }} \cr \nonumber\\ 
&&\left \{\sin (\theta_{q_1} + \theta_{q_2}) \left [ \frac{1}{2\pi i}\int_{-\infty}^{\infty} dp_0 e^{-ip_o(t_1 - t_n)} \right . \right . \cr \nonumber\\
&& \left. \left . \frac{\cos (\theta_{q_2} + \theta_{q_3})\cos (\theta_{q_3} + \theta_{q_4}) \cdots \cos (\theta_{q_{j-1}} + \theta_{q_{j}})}{(p_0 - 2\Lambda _{q_2} + i\varepsilon) (p_0 - 2\Lambda _{q_3} + i\varepsilon) \cdots (p_0 - 2\Lambda _{q_{j-1}} + i\varepsilon)(p_0 -2 \Lambda _{q_j} + i\varepsilon)} \right ] \right . \cr \nonumber\\
&& \left . \sin (\theta_{q_{j}} + \theta_{q_{j+1}}) \left [ \frac{1}{2\pi i}\int_{-\infty}^{\infty} dk_0 e^{-ik_o(t_1 - t_n)} \right . \right . \cr \nonumber\\
&&\left. \left. \frac{ \cos (\theta_{q_{1}} + \theta_{q_{n}}) \cos (\theta_{q_{n}} + \theta_{q_{n-1}}) \cdots \cos (\theta_{q_{j+2}} + \theta_{q_{j+1}})}{(k_0 - 2\Lambda _{q_1} + i\varepsilon)(k_0 - 2\Lambda _{q_n} + i\varepsilon) \cdots (k_0 - 2\Lambda _{q_{j+2}} + i\varepsilon)(k_0 -2 \Lambda _{q_{j+1}} + i\varepsilon)} \right ] \right \} . \label{n_ord_cumulant_convolution_SD}
\end{eqnarray}

We can now proceed, obtain the remaining integrals in (\ref{n_ord_cumulant_convolution_SD}), and obtain the $n$-order term of perturbative expansion, given by (\ref{nth_ord_cum2}) , (\ref{nth_ord_cum4}) and (\ref{nth_ord_cum}), hence confirming these results. However, much more importantly, the structure of (\ref{n_ord_cumulant_convolution_SD}) allows us to directly resum the terms of the perturbative expansion (\ref{G_cumulant_exp}). One immediately notices that in each term of the sum over $j$ in (\ref{n_ord_cumulant_convolution_SD}), there is a product of two terms of two equations for the self-consistent resummed retarded function $\tilde{G}_{R, q, q^\prime }(p_0)$. The form of the equation is
\begin{eqnarray} 
& & G_{R, q}(p_0) + G_{R, q}(p_0) \sum_{q^\prime \in \Gamma ^{\pm }} \frac{-i 2 \delta h}{N} \cos (\theta_{q} + \theta_{q^\prime}) G_{R, q^\prime}(p_0) \cr \nonumber \\ 
& & + G_{R, q}(p_0) \sum_{q^\prime, q^{\prime \prime} \in \Gamma ^{\pm }} \frac{-i2 \delta h}{N} \cos (\theta_{q} + \theta_{q^\prime}) G_{R, q^\prime}(p_o) \frac{-i2 \delta h}{N} \cos (\theta_{q^\prime } + \theta_{q^{\prime \prime}}) G_{R, q^{\prime \prime}}(p_0) + \dots \cr \nonumber \\
& & = G_{R, q}(p_0) + G_{R, q}(p_0) \sum_{q^\prime \in \Gamma ^{\pm }} \frac{-i2 \delta h}{N} \cos (\theta_{q} + \theta_{q^\prime}) \tilde{G}_{R, q^\prime , q^{\prime \prime} }(p_0) . \label{self_consistent_resummed}
\end{eqnarray}

Here, $G_{R,q}(p_0)$ is the retarded function (\ref{WT_Green_2point_retarded}). By resummation of all terms in (\ref{self_consistent_resummed}), one obtains function $\tilde{G}_{R, q, q^\prime }(p_0)$:
\begin{equation}
\tilde{G}_{R, q, q^\prime}(p_0) =G_{R,q}(p_0)\frac{1}{1 - \frac{-i2 \delta h}{N} \cos (\theta_{q} + \theta_{q^\prime}) G_{R, q^\prime}(p_0) } = G_{R,q}(p_0) R_{q, q^\prime}({\bf A}(p_0, \delta h , N)) . \label{G_resummed}
\end{equation}

The sum of infinite geometric series corresponds here to a matrix function 
 ${\bf R}({\bf A} (p_0, \delta h , N)) = (1 - ({\bf A} (p_0, \delta h , N)) )^{-1}$ of an $N \times N$ matrix ${\bf A} (p_0, \delta h , N)$. $R_{q, q^\prime}({\bf A}(p_0, \delta h , N))$ denotes matrix elements of ${\bf R}({\bf A} (p_0, \delta h , N))$. 
Elements of the matrix ${\bf A} (p_0, \delta h , N)$ are
\begin{equation}
A_{q_1, q_2} (p_0, \delta h , N) = \frac{-i2 \delta h}{N} \cos (\theta_{q_1} + \theta_{q_2}) G_{R, q_2}(p_0) = \frac{\frac{2 \delta h}{N} \cos (\theta_{q_1} + \theta_{q_2})} {(p_0 - 2\Lambda _{q_2} + i\varepsilon)} , \qquad q_1, q_2 \in \Gamma ^{\pm } . \label{mat_A} 
\end{equation}

They are numerated by fermion momenta $q_1$ and $q_2$ as indices. As explained in Section~\ref{perturbative_calculations}, momenta $q$ and energy $\epsilon (q)$ of a Bogoliubov fermion is not conserved by the perturbations (\ref{V_Bogoliubov}) and (\ref{V_matrix}). Thus, (\ref{self_consistent_resummed}) can be conditionally called a generalized Schwinger--Dyson equation.

Perturbative expansion (\ref{G_cumulant_exp}) is then resummed using (\ref{1th_ord_cluster}), (\ref{n_ord_cumulant_convolution_SD}), (\ref{self_consistent_resummed}), (\ref{G_resummed}), and (\ref{mat_A}). We~obtain
\begin{eqnarray}
&& \log[\mathcal{G}(t)] = \sum_{n=1}^\infty\frac{(-i)^n}{n!}\int_0^t dt_1 \dots \int_0^t dt_n \bra{g_0}T[V_I(t_1) \dots V_I(t_n)]\ket{g_0}_C \cr \nonumber \\ 
&& = i \frac{\delta h}{N}\left ( \sum_{q \in \Gamma \pm } \cos 2\theta _q \right ) t\cr \nonumber \\
&& - \frac{2 (\delta h)^2}{N^2} \int_{0}^t dt_1 \int_{0}^t dt_n \theta(t_1 - t_n) \sum_{q_1, q_2, q_3, q_4 \in \Gamma ^{\pm }} \left \{\sin (\theta_{q_1} + \theta_{q_2}) \sin (\theta_{q_{3}} + \theta_{q_{4}}) \right . \cr \nonumber\\ 
&& \left. \left [ \frac{1}{2\pi i}\int_{-\infty}^{\infty} dp_0 e^{-ip_o(t_1 - t_n)} \frac{1}{(p_0 - 2\Lambda _{q_2} + i\varepsilon)} R_{q_2, q_3}({\bf A}(p_0, \delta h , N)) \right ] \right . \cr \nonumber\\
&& \left . \left [ \frac{1}{2\pi i}\int_{-\infty}^{\infty} dk_0 e^{-ik_o(t_1 - t_n)} \frac{1}{(k_0 - 2\Lambda _{q_1} + i\varepsilon)} R_{q_1, q_4}({\bf A}(k_0, \delta h , N)) \right ] \right \} . \label{G_cumulant_exp_resummed} 
\end{eqnarray}

Expression (\ref{G_cumulant_exp_resummed}) can be evaluated by a combination of analytical and numerical methods, as described in the next section.

Finally, we make an important remark regarding the matrix ${\bf A} (p_0, \delta h , N)$ and two assumptions about matrix function ${\bf R}({\bf A} (p_0, \delta h, N))$: 
\begin{itemize}
\item[+] {\bf Remark:} 
 It is hard to see how matrix ${\bf A} (p_0, \delta h , N)$, given by (\ref{mat_A}), can be diagonalizable for all values of $p_0$. However, for the purpose of evaluating (\ref{G_cumulant_exp_resummed}), it is important that matrix $\mathbf {\hat{A}} (\delta h , N)$, which comprises only of nonsingular ``on shell'' elements of the matrix ${\bf A} (p_0, \delta h , N)$,
\begin{equation}
\hat{A}_{q_1, q_2} (\delta h , N) = \left \{ \begin{array}{l@{\,,\qquad }l} \frac{\frac{2 \delta h}{N} \cos (\theta_{q_1} + \theta_{q_2})} {( 2\Lambda _{q_1} - 2\Lambda _{q_2} + i\varepsilon)} & q_1, q_2 \in \Gamma ^{\pm }, 2\Lambda _{q_1} \neq 2\Lambda _{q_2} \\
0 & q_1, q_2 \in \Gamma ^{\pm }, 2\Lambda _{q_1} = 2\Lambda _{q_2} \end{array} \right . , \label{mat_A_q}
\end{equation}
is diagonalizable. Matrix $\mathbf {\hat{A}} ( \delta h , N)$, defined by (\ref{mat_A_q}), is skew-Hermitian and, therefore, it is diagonalizable by a unitary transformation $\mathbf{U}^{\dagger}\mathbf {\hat{A}} ( \delta h , N) \mathbf{U}$. Thus, in the case of the ``on shell'' matrix $\mathbf {\hat{A}} (\delta h , N)$, matrix function $\mathbf {R}(\mathbf {\hat{A}} ( \delta h , N))$ corresponds to a sum of infinite geometric series of a diagonalizable matrix $\mathbf {\hat{A}} ( \delta h , N)$.

\item[+] {\bf Assumption 1:} \label{asumption_1} We assume that, if all eigenvalues of the matrix $\mathbf {\hat{A}} ( \delta h , N)$ are within the radius of convergence $|z|<1$ of infinite geometric series $1 + z + z^2 + \dots = 1/(1 - z)$, the validity of (\ref{self_consistent_resummed}), (\ref{G_resummed}), and (\ref{G_cumulant_exp_resummed}) can be extended outside the radius of convergence by methods of analytic continuation. After integrations over $p_0$ and $k_0$ in (\ref{G_cumulant_exp_resummed}), parts that contain only nonsingular ``on shell'' elements defining the matrix $\mathbf {\hat{A}} ( \delta h , N)$ are resummed separately. 

\item[+] {\bf Assumption 2:} \label{asumption_2} Parts of (\ref{G_cumulant_exp_resummed}) that contain contributions from singular ``on shell'' elements of the matrix ${\bf A} (p_0, \delta h , N)$ are treated by assuming that singular points are also inherited by its matrix function ${\bf R}({\bf A} (p_0, \delta h, N))$. This is certainly true for all poles of finite order, because they are within the radius of convergence of the nonsingular ``on shell'' part $\mathbf {R}(\mathbf {\hat{A}} ( \delta h , N))$. Proof is given in Appendix \ref{Poles_R_A}. The same procedure is followed in integration over $k_0$.
\end{itemize}

Correctness of these assumptions and the resummation procedure based on them is checked in the next section. There, results of the resummation of perturbative expansion (\ref{G_cumulant_exp}) of LE $\mathcal{L}(t)= |\mathcal{G}(t)|^2$ are compared with the exact numerical results of diagonalization of the Hamiltonian (\ref{XY_Hamiltonian}) and evaluation of (\ref{complex_amplitude_LE}).

\section{Results} \label{results}
Following the above assumption, we first integrate over $p_0$ and $k_0$ in (\ref{G_cumulant_exp_resummed}) by closing the contours in the lower complex semiplane and collecting contributions from all residues. Then, we take the time integrals in (\ref{G_cumulant_exp_resummed}) and obtain the final resummed form of (\ref{G_cumulant_exp}):
\vspace{-6pt}
\begin{eqnarray}
&& \log[\mathcal{G}(t)] = i \frac{\delta h}{N}\left ( \sum_{q \in \Gamma \pm } \cos 2\theta _q \right ) t - \frac{2 (\delta h)^2}{N^2} \sum_{q_1, q_2, q_3, q_4 \in \Gamma ^{\pm }} \left \{\sin (\theta_{q_1} + \theta_{q_2}) \sin (\theta_{q_{3}} + \theta_{q_{4}}) \right . \cr \nonumber\\ 
&& \left. \times \left [ \left (R_{q_2, q_3}({\bf \hat {A}}(\delta h , N)) + \sum_{n=2}^{\infty} R_{q_2; q_2, q_3}({\bf A}( 2\Lambda _{q_2}, \delta h , N))_{(n-1) \ \textrm{amp.}} \frac{1}{(n-1)!}\frac{d^{n-1}}{(d 2\Lambda _{q_2})^{n-1}} \right) \right. \right . \cr \nonumber \\ 
&& \left. \left . \times \left (R_{q_1, q_4}({\bf \hat {A}}(\delta h , N)) + \sum_{m=2}^{\infty}R_{q_1; q_1, q_4}({\bf A}( 2\Lambda _{q_1}, \delta h , N))_{(m-1) \ \textrm{amp.}} \frac{1}{(m-1)!}\frac{d^{m-1}}{(d 2\Lambda _{q_1})^{m-1}} \right ) \right . \right . \cr \nonumber\\
&& \left . \left . \times \mathcal{F} ( 2\Lambda _{q_1}, 2\Lambda _{q_2}, t) \right . \right. \cr\nonumber \\
&& \left . \left . + 2 \left (R_{q_2, q_3}({\bf \hat {A}}(\delta h , N)) + \sum_{n=2}^{\infty} R_{q_2; q_2, q_3}({\bf A}( 2\Lambda _{q_2}, \delta h , N))_{(n-1) \ \textrm{amp.}} \frac{1}{(n-1)!}\frac{d^{n-1}}{(d 2\Lambda _{q_2})^{n-1}} \right) \right . \right . \cr \nonumber \\
&& \left . \left . \times \left ( \sum_{q_6 \in \Gamma ^{\pm }} ^{\Lambda _{q_6} \ne \Lambda _{q_1} } \sum_{m=1}^{\infty}R_{q_1; q_6, q_4}({\bf A}( 2\Lambda _{q_6}, \delta h , N))_{m \ \textrm{amp.}} \frac{1}{(m-1)!}\frac{d^{m-1}}{(d 2\Lambda _{q_6})^{m-1}} \right ) \frac{\mathcal{F} ( 2\Lambda _{q_6}, 2\Lambda _{q_2}, t) }{ 2\Lambda _{q_6} - 2\Lambda _{q_1}} \right . \right . \cr \nonumber \\
&& \left . \left . + \left ( \sum_{q_5, q_6 \in \Gamma ^{\pm }} ^{\Lambda _{q_5} \ne \Lambda _{q_2}, \Lambda _{q_6} \ne \Lambda _{q_1} } \sum_{n=1}^{\infty}R_{q_2; q_5, q_3}({\bf A}( 2\Lambda _{q_5}, \delta h , N))_{n \ \textrm{amp.}} \frac{1}{(n-1)!}\frac{d^{n-1}}{(d 2\Lambda _{q_5})^{n-1}} \right . \right. \right. \cr \nonumber\\
&& \left. \left . \left . \times \sum_{m=1}^{\infty}R_{q_1; q_6, q_4}({\bf A}( 2\Lambda _{q_6}, \delta h , N))_{m \ \textrm{amp.}} \frac{1}{(m-1)!}\frac{d^{m-1}}{(d 2\Lambda _{q_6})^{m-1}}\right ) \right . \right . \cr\nonumber \\
&& \left. \left . \times \frac{\mathcal{F} ( 2\Lambda _{q_6}, 2\Lambda _{q_5}, t) }{ (2\Lambda _{q_6} - 2\Lambda _{q_1})(2\Lambda _{q_5} - 2\Lambda _{q_2})} \right ] \right \} . \label{G_cumulant_exp_resummed_int_total} 
\end{eqnarray}

Here, an object with three indices $R_{q_1, q_2, q_3}({\bf A}( 2\Lambda _{q_2}, \delta h , N))_{n \ \textrm{amp.}}$ is a part of matrix element of infinite geometric series containing only $n$-th order singular terms with ``amputated'' legs, which renders them nonsingular. Its meaning and how it appears through application of the residue theorem is explained in Appendix \ref{Poles_R_A}. Function $ \mathcal{F} (x, y, t) $ in (\ref{G_cumulant_exp_resummed_int_total}) is a result of time integrations in (\ref{G_cumulant_exp_resummed}), and it is given by 
\begin{equation}
\mathcal{F} (x, y, t) = \frac{1-e^{-i(x + y)t}}{(x+y)^2} - \frac{it}{x+y} .
\end{equation}

Derivatives in (\ref{G_cumulant_exp_resummed_int_total}) are over all variables $x$ and $y$ of the function $\mathcal{F} (x, y, t)$, which appear to the right of the derivative signs. In the large time limit $t \gg 1 /(x+y)$, leading term of $n$-th derivative of $\mathcal{F} (x, y, t)$ over $x$ or $y$, is equal to $((-1)^{n+1}(it)^n /n!) e^{-i(x + y)t}/(x+y)^2$. Hence, to obtain a convergent expression in this limit, summation in (\ref{G_cumulant_exp_resummed_int_total}) should be carried out over derivatives of all order, multiplied by ``amputated'' singular part of geometric~series. 

However, by comparing matrix elements of ${\bf A}(p_0, \delta h , N)$ and ${\bf \hat {A}}(\delta h , N)$, in (\ref{mat_A}) and (\ref{mat_A_q}), and the definition of ``amputated'' $n$-th order singular part of geometric series, in (\ref{AppB_Res4}) and (\ref{AppB_Res6}), with respect to simple pole part in (\ref{AppB_Res5}), we notice that

\begin{equation}
\left | { R_{q_1, q_2, q_3}({\bf A}( 2\Lambda _{q_2}, \delta h , N))_{n \ \textrm{amp.}} \over R_{q_1, q_2, q_3}({\bf A}(2\Lambda _{q_2}, \delta h , N))_{1 \ \textrm{amp.}}} \right | \le \left(\frac{2\delta h}{N} \right )^{n-1} \left (N\ \mathrm{max} \left |{\bf R}({\bf \hat {A}}(\delta h , N)) \right | \right )^{n-1} \propto \left (2\delta h \right )^{n-1} . \label{G_cumulant_exp_resummed_spa}
\end{equation}

As a consequence of this, in a small perturbation limit $2\delta h \ll 1$, the contribution of ``amputated'' simple pole parts of matrix elements of ${\bf R}({\bf A} (p_0, \delta h, N))$ to the resummed perturbative expansion (\ref{G_cumulant_exp_resummed_int_total}), dominates over the contribution of higher order ``amputated'' singular parts. By considering only the simple pole part contribution to (\ref{G_cumulant_exp_resummed_int_total}), we obtain a much simpler expression:
\vspace{-9pt}
\begin{eqnarray}
&& \log[\mathcal{G}(t)] = i \frac{\delta h}{N}\left ( \sum_{q \in \Gamma \pm } \cos 2\theta _q \right ) t - \frac{2 (\delta h)^2}{N^2} \sum_{q_1, q_2, q_3, q_4 \in \Gamma ^{\pm }} \left \{\sin (\theta_{q_1} + \theta_{q_2}) \sin (\theta_{q_{3}} + \theta_{q_{4}}) \right . \cr \nonumber\\ 
&& \left. \times \left [ R_{q_2, q_3}({\bf \hat {A}}(\delta h , N)) R_{q_1, q_4}({\bf \hat {A}}(\delta h , N)) \mathcal{F} ( 2\Lambda _{q_1}, 2\Lambda _{q_2}, t) + \left ( 2 R_{q_2, q_3}({\bf \hat {A}}(\delta h , N)) \right. \right. \right . \cr \nonumber
&& \left. \left . \left . \times \frac{2\delta h}{N}\sum_{q_7, q_8 \in \Gamma ^{\pm }} ^{\Lambda _{q_8} \ne \Lambda _{q_1} } \frac{R_{q_1; q_8, q_7}({\bf \hat{A}}_{q_1}(\delta h , N))\cos (\theta_{q_7} + \theta_{q_8}) R_{q_8, q_4}({\bf \hat {A}}(\delta h , N)) }{ 2\Lambda _{q_8} - 2\Lambda _{q_1}} \mathcal{F} ( 2\Lambda _{q_8}, 2\Lambda _{q_2}, t) \right ) \right . \right . \cr \nonumber\\
&& \left . \left . + \left (\frac{2\delta h}{N}\right )^2 \sum_{q_6, q_5, q_7, q_8 \in \Gamma ^{\pm }} ^{\Lambda _{q_6} \ne \Lambda _{q_2}, \Lambda _{q_8} \ne \Lambda _{q_1} } \left ( \frac{R_{q_2; q_6, q_5}({\bf \hat {A}}_{q_2}(\delta h , N))\cos (\theta_{q_5} + \theta_{q_6}) R_{q_6, q_3}({\bf \hat {A}}(\delta h , N)) }{ 2\Lambda _{q_6}- 2\Lambda _{q_2} } \right . \right. \right. \cr \nonumber\\
&& \left. \left . \left . \times \frac{R_{q_1; q_8, q_7}({\bf \hat{A}}_{q_1}(\delta h , N))\cos (\theta_{q_7} + \theta_{q_8}) R_{q_8, q_4}({\bf \hat{A}}(\delta h , N)) }{ 2\Lambda _{q_8} - 2\Lambda _{q_1}} \mathcal{F} ( 2\Lambda _{q_8}, 2\Lambda _{q_6}, t) \right ) \right ] \right \} . \label{G_cumulant_exp_resummed_int} 
\end{eqnarray}

As we explain in Appendix~\ref{Poles_R_A}, an object with three indices $R_{q_1; q_2, q_3}({\bf \hat{A}}_{q_1}(\delta h , N)$ is a matrix element of infinite geometric series of the matrix ${\bf \hat{A}}_{q_1}(\delta h , N)$ from the set of matrices $\{{\bf \hat{A}}_{q_1}(\delta h , N): q_1 \in \Gamma ^{\pm }\} $, denoted by the index $q_1$. Each of these matrices is defined by its~elements, 
\begin{equation}
\hat{A}_{q_1; q_2, q_3} (\delta h , N) = \left \{ \begin{array}{l@{\,,\qquad }l} \frac{\frac{2 \delta h}{N} \cos (\theta_{q_1} + \theta_{q_3})} {( 2\Lambda _{q_2} - 2\Lambda _{q_3} + i\varepsilon)} & q_2, q_3 \in \Gamma ^{\pm }, 2\Lambda _{q_2} \neq 2\Lambda _{q_3} \\
0 & q_2, q_3 \in \Gamma ^{\pm }, 2\Lambda _{q_2} = 2\Lambda _{q_3} \end{array} \right . . \label{mat_A_q_q_q}
\end{equation}

Diagonalizability of the set of matrices $\{{\bf \hat{A}}_{q_1}(\delta h , N): q_1 \in \Gamma ^{\pm }\} $ is checked numerically. 

\begin{figure}[!htb] 
\includegraphics{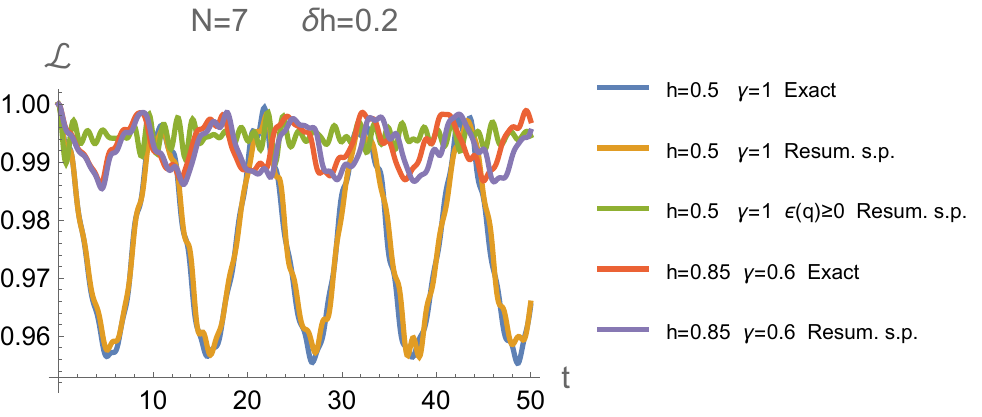}
\caption{\small{LE for the local quench in the AFM phase of topologically frustrated Ising and XY chains. Results of resummation of perturbative expansion of LE show close correspondence with the exact numerical result. Due to application of the simple pole contribution approximation (\ref{G_cumulant_exp_resummed_int}) to the full resummation result (\ref{G_cumulant_exp_resummed_int_total}), small difference can be noticed to grow in the large time limit. Perturbation $\delta h =0.2$ is roughly equal to the size of the gap $\Delta$ between the nondegenerate ground state and excited states (in this case, $\Delta \approx 0.18$ and $\Delta \approx 0.22$ for Ising and XY chain, respectively). In both cases, eigenvalues of diagonalizable matrices $\mathbf {\hat{A}} ( \delta h , N)$ and $\{{\bf \hat{A}}_{q_1}(\delta h , N): q_1 \in \Gamma ^{\pm }\} $ are within the radius of convergence of the geometric series, as Assumptions 1 and 2 require. For comparison, resummed LE with a negative energy mode taken to be nonnegative is illustrated for Ising chain (green line on the plot). As a result of PBC, Bogoliubov fermion energy modes $q=0, \pi$ can be negative (see (\ref{en_eps}) and (\ref{excitation_energies});  it is $q=0$ mode that is negative for the parameters chosen on the plot). To be precise, taking it to be nonnegative means that there are no exceptions to (\ref{en_eps}). As discussed in Section  \ref{perturbative_calculations}, in the thermodynamic limit,  such an approach  eventually would correspond to the procedure applied in \cite{Silva1}.  }} \label{fig2}
\end{figure}

Results of resummation of perturbative expansion of LE in the approximation $2\delta h \ll 1$, given by (\ref{G_cumulant_exp_resummed_int}), are compared with the exact numerical results in Figure \ref{fig2} for Ising and XY chain. Strength of the perturbation $\delta h$ is chosen so that it is approximately equal to the size of the gap between the nondegenerate ground state and excited states. For the parameters chosen (AFM phase of topologically frustrated Ising and XY chains), the gap closes as $\propto 1/N^2$ (see Section \ref{perturbative_calculations} and references \cite{groupLE, Dong1, Dong2, group1, group2, group3, group4, group5, group6, group7, Odavic1} for details). As long as the ground state of the chain is nondegenerate and the condition $2\delta h \ll 1$ applies, approximation (\ref{G_cumulant_exp_resummed_int}) very closely corresponds to the exact results for LE. Beyond $2\delta h \ll 1$ limit, a full resummation result (\ref{G_cumulant_exp_resummed_int_total}) is necessary to achieve correspondence with the exact result. Results of resummation of perturbative expansion of LE confirm in this way the validity of analyticity Assumptions 1 and 2 (introduced at the end of Section \ref{projected_functions_resummation}) on which it was based. 

Although the results shown in Figure \ref{fig2}  are for relatively small systems ($N=7$), distinctive LE oscillations of topologically frustrated AFM chains persist with the system size and are strongly dependent on it, both in amplitude (with an asymptotic value for a large size), and in the characteristic period of revivals, which is proportional to $N^2$. Results for larger systems were presented and discussed in \cite{groupLE}. This behavior of LE is quantitatively and qualitatively very different from the behavior for unfrustrated chains, and this is ultimately related to closing of the energy gap in the AFM phase of topologically frustrated chains in the thermodynamic limit, while it remains gapped for unfrustrated~chains.

\section{Conclusions} \label{conclusions}

As we have seen in this work, within the framework of FTPFT, it is possible to carry out a full resummation of the perturbative expansion of LE for a local quench of Ising and XY spin chains, where transverse magnetic field is suddenly perturbed at a single spin site. Resummation was achieved by applying an FTPFT formalism of WTs of projected functions and their convolution products \cite{Dadic4}. In this way, a resummation of generalized Schwinger--Dyson equations for the two-point retarded functions in the basic LE ``bubble'' diagrams was carried out first, leading directly to a resummation of the perturbative expansion of LE. Resummation was possible under two simple and quite general analyticity assumptions introduced at the end of Section \ref{projected_functions_resummation}. Validity of these assumptions was tested and confirmed by a close correspondence of the results of resummation of perturbative expansion of LE and the exact numerical results. We have also obtained a somewhat simplified expression for the resummed LE, which approximates it quite well in the small perturbation limit.

Further development, implementing a quasi-particle fermion excitation approximation with shifted energy poles of two-point retarded functions, will be the subject of future work, with the purpose to extend the results presented here beyond the region of validity of the introduced analyticity assumptions. We expect that this and further work on the subject should open a possibility to apply the approach also to other spin chains that are integrable via mapping to second quantized noninteracting fermions, and for different types and strengths of perturbation that breaks the translational symmetry of these models. The approach could also be applied to simulate the presence of disorder and interactions, as in different many-body localization regimes \cite{Nadkishore1}. Of close interest are also possible concepts for quantum technology, like energy storage \cite{Catalano1}, or information storage. 

Given that FTPFT does not depend on the assumption of adiabatic switching on the perturbation, we consider it a promising theory when considering finite time behavior for systems with degenerate ground states breaking the symmetry of the model. Standard applications of field theory and Gell-Mann--Low theorem in linking interacting and noninteracting Green's functions assume adiabatic switching on and a nondegenerate ground state (vacuum) \cite{Coleman}. While we have explicitly avoided the former assumption by applying a sudden quench scenario, the latter assumption is still adhered to in this paper. Hopefully, further work will shed more light on whether this latter assumption is really necessary in FTPFT applications like the one presented here.

\vspace{6pt}

\acknowledgments{D.K. wishes to express particular thanks to Ivan Dadi\'{c}, Vanja Mari\'{c}, Gianpaolo Torre, Fabio Franchini and Salvatore Marco Giampaolo for discussions and motivation.}  

\appendix
\section{Diagonalization 
 of Hamiltonian} \label{diagonalization}

Diagonalization of the Hamiltonian (\ref{XY_Hamiltonian}) is performed by applying JWT \cite{JWT} to the problem to spinless fermions, followed by a discrete Fourier transformation and a Bogoliubov transformation in momentum space. A feature of JWT employed here is that one can choose it to map spin ups as empty and spin downs as occupied fermionic states, or vice versa. Exploiting this feature of JWT, one can check that after a subsequent Fourier and Bogoliubov transformation, the ground state of diagonalized Hamiltonian can be expressed as a vacuum of Bogoliubov fermions in momentum space. To achieve this, for $N$ even and $h \ne 0$, and also for $N$ odd and $h > 0$, one chooses the following JWT to fermionic operators: 
\begin{eqnarray}
c_j = \bigotimes_{k=1}^{j-1} \sigma_k^{z} \sigma_j^{-}, \qquad \qquad c_j^\dagger = \bigotimes_{k=1}^{j-1} \sigma_k^{z} \sigma_j^{+} . \label{JWT2}
\end{eqnarray}
where $\sigma_j^{\pm } = (\sigma_j^x \pm i \sigma_j^y)/2$ are the Pauli spin ladder operators. With the same purpose, for $N$ odd and $h < 0$, we employ a redefined JWT: 
\begin{eqnarray}
c_j = \bigotimes_{k=1}^{j-1} \sigma_k^{z} \sigma_j^{+}, \qquad \qquad c_j^\dagger = \bigotimes_{k=1}^{j-1} \sigma_k^{z} \sigma_j^{-} , \label{JWT1}
\end{eqnarray}

After JWT, either (\ref{JWT2}) or (\ref{JWT1}), we apply a discrete Fourier transform to momentum space fermionic operators, 
\begin{eqnarray}
c_q = \frac{1}{\sqrt{N}}\sum_{j=1}^{N}e^{-iqj}c_j , \qquad \qquad c_q^\dagger = \frac{1}{\sqrt{N}}\sum_{j=1}^{N}e^{iqj}c_j^\dagger , \label{c_DFT}
\end{eqnarray}
with momenta $\{q = 2\pi k /N : k =0, 1, \dots, N - 1 \} = \Gamma^- $ or $\{q = 2\pi(k + \frac{1}{2})/N : k =0, 1, \dots, N - 1 \} = \Gamma^+$. Finally, a (unitary) Bogoliubov transformation is applied: 
\begin{eqnarray}
c_q = \cos \theta_q \eta _q - i\sin \theta_q \eta _{-q}^\dagger \qquad \qquad c_{-q}^\dagger = -i \sin \theta_q \eta _q + \cos \theta_q \eta _{-q}^\dagger . \label{BT}
\end{eqnarray}

In this way, by choosing JWT (\ref{JWT2}) for $N$ even and $h \ne 0$, and also for $N$ odd and $h > 0$, and JWT (\ref{JWT1}) for $N$ odd and $h < 0$, and then applying (\ref{c_DFT}) and (\ref{BT}), one obtains that: 
\begin{itemize}
\item[+] The Hamiltonian is divided in two parity sectors corresponding to eigenspaces of parity operator $\Pi^z = \bigotimes_{j=1}^N \sigma_j^z$ with eigenvalues equal to $+1$ or $-1$,
\begin{equation}
H = \frac{1+\Pi^z}{2}H^+ \frac{1+\Pi^z}{2} + \frac{1 - \Pi^z}{2}H^- \frac{1 -\Pi^z}{2} . \label{diagonalized_Hamiltonian}
\end{equation}
\item[+] In each parity sector, the Hamiltonian is diagonal in terms of Bogoliubov fermions and their creation $\eta _{q}^\dagger$ and annihilation operators $\eta _{q}$,
\begin{equation}
H^{\pm } = \sum_{q \in \Gamma ^{\pm }} \epsilon (q) \left (\eta _{q}^\dagger \eta _{q} - \frac{1}{2}\right) ,
\end{equation}
with excitation energies of Bogoliubov fermions $\epsilon (q) = 2\Lambda_q$ given by (\ref{en_eps}) and (\ref{excitation_energies}). 
\item[+] Nondegenerate ground state of (\ref{diagonalized_Hamiltonian}) is represented as a Bogoliubov vacuum. For $N$ even and $h \ne 0$, and also for $N$ odd and $h < 0$, ground state $\ket{g_0^+}$ is a state of positive parity $\Pi^z \ket{g_0^+} = \ket{g_0^+}$. It is a Bogliubov vacuum, meaning that 
\begin{equation}
\eta _{q}\ket{g_0^+} = 0, \qquad \forall q \in \Gamma ^+ . \label{gs_+_Bogoliubov}
\end{equation}

For $N$ odd and $h > 0$, the ground state is a state $\ket{g_0^{-}}$ of negative parity $\Pi^z \ket{g_0^-} = -\ket{g_0^-}$. It is also a Bogoliubov vacuum, 
\begin{equation}
\eta _{q}\ket{g_0^-} = 0, \qquad \forall q \in \Gamma ^- . \label{gs_-_Bogoliubov}
\end{equation}
\end{itemize}

\section{Poles and Residues of the Matrix Function \boldmath{${\bf R}({\bf A} (p_0, \delta h, N))$}} \label{Poles_R_A}

Here, we use the fact that geometric series ${\bf R}({\bf A} (p_0, \delta h, N))$ has, within its radius of convergence, an inverse matrix ${\bf R}^{-1}({\bf A} (p_0, \delta h, N)) = {\bf I} - {\bf A} (p_0, \delta h, N)) $, since
\begin{equation}
{\bf R}({\bf A}) {\bf R}^{-1}({\bf A}) = ({\bf I} + {\bf A} + {\bf A}^2 + \dots )({\bf I} - {\bf A} ) = {\bf I} . \label{R_A_inverse}
\end{equation}

Symbols for the complex variable $p_0$, and parameters $\delta h$ and $N$ of the matrix ${\bf A} (p_0, \delta h, N))$ are suppressed here for brevity. Taking the $p_0$ derivative of (\ref{R_A_inverse}), one obtains
\begin{equation}
\frac{d \left [{\bf R}({\bf A}) {\bf R}^{-1}({\bf A})\right] }{dp_0}=\frac{d {\bf R}({\bf A}) }{dp_0} ({\bf I} - {\bf A} ) - {\bf R}({\bf A})\frac{d {\bf A}}{dp_0} = 0 . \label{R_A_inverse_ident_der}
\end{equation}

Then, multiplying (\ref{R_A_inverse_ident_der}) with ${\bf R}({\bf A} (p_0, \delta h, N))$  and  using (\ref{R_A_inverse})  we obtain
\begin{equation}
\frac{d {\bf R}({\bf A}) }{dp_0} = {\bf R}({\bf A})\frac{d {\bf A} }{dp_0} {\bf R}({\bf A}) . \label{R_A_derivative}
\end{equation}

In (\ref{mat_A}), we observe that matrix ${\bf A} (p_0, \delta h, N))$ contains only poles of first order with respect to the complex variable $p_0$. By isolating parts of matrix elements of ${\bf R}({\bf A} (p_0, \delta h, N))$ with these simple poles, it is possible to calculate the sum of their residues at these singular points, with the help of relation
\begin{eqnarray}
&& \sum_{q_2 \in \Gamma \pm } \textrm{Res} \ R_{q_1; q_2, q_4}({\bf A} (2\Lambda _{q_2}, \delta h, N))_{\textrm{sim. pole}} \cr \nonumber \\
&& = -\sum_{q_2 \in \Gamma ^ \pm } \lim_{p_0 \rightarrow 2\Lambda _{q_2} } \left [(p_0 - 2\Lambda _{q_2} )^2\frac{d R_{q_1, q_4}({\bf A}(p_0, \delta h , N))_{\textrm{sim. pole}} }{dp_0} \right ] \cr \nonumber \\
&& = - \sum_{q_2, q_3 \in \Gamma ^\pm }\lim_{p_0 \rightarrow 2\Lambda _{q_2} } \left [(p_0 - 2\Lambda _{q_2} )^2 \right . \cr \nonumber \\
&& \left. \times R_{q_1, q_3}({\bf \hat {A}}_{q_1}(p_0, \delta h , N)) ) \frac{d A_{q_3, q_2} (p_0, \delta h, N))}{dp_0} R_{q_2, q_4 }({\bf \hat {A}}(p_0, \delta h , N)) ) \right ] \cr \nonumber \\
&& = \sum_{q_2, q_3 \in \Gamma ^\pm } R_{q_1; q_2, q_3}({\bf \hat {A}}_{q_1}(\delta h , N)) )\cos (\theta_{q_3} + \theta_{q_2}) R_{q_2, q_4 }({\bf \hat {A}}(\delta h , N)) ) . \label{AppB_Res1}
\end{eqnarray}

The derivative in this relation is calculated with the help of (\ref{R_A_derivative}). Isolation of simple pole terms from higher order singular terms in the Laurent expansion is accomplished using the nonsingular ``on shell'' part $\mathbf {\hat{A}} (\delta h , N))$, with the elements (\ref{mat_A_q}), and the set of matrices $\{{\bf \hat{A}}_{q_1}(\delta h , N): q_1 \in \Gamma ^{\pm }\} $, with the elements (\ref{mat_A_q_q_q}). $R_{q_1; q_2, q_3}({\bf \hat{A}}_{q_1}(\delta h , N)$ is a matrix element of infinite geometric series of the matrix ${\bf \hat{A}}_{q_1}(\delta h , N)$.

By applying the standard procedure, one can show, using \ref{R_A_derivative} in a similar fashion, that the sum of residues of all $n$-th order singular terms in the Laurent expansion ($n \ge 2$) are equal to zero,\vspace{-9pt}
\begin{eqnarray}
&& \sum_{q_2 \in \Gamma \pm } \textrm{Res} \ R_{q_1; q_2, q_3}({\bf A} (2\Lambda _{q_2}, \delta h, N))_{n \ \textrm{sing.}} \cr \nonumber \\
&& = \sum_{q_2 \in \Gamma \pm } \frac{1}{(n-1)!} \lim_{p_0 \rightarrow 2\Lambda _{q_2} }\frac{d^{n-1}}{dp_0^{n-1}} \left [(p_0 - 2\Lambda _{q_2} )^n R_{q_1, q_3}({\bf A}(p_0, \delta h , N))_{n \ \textrm{sing.}} \right ] = 0 . \label{AppB_Res2}
\end{eqnarray}

However, if $R_{q_1; q_2, q_3}({\bf A} (p_0, \delta h, N))_{n \ \textrm{sing.}}$ are multiplied by some function of $p_0$ (for example, $e^{-ip_0 t}$ as in (\ref{G_cumulant_exp_resummed})), we end up with\vspace{-8pt}
\begin{eqnarray}
&& \sum_{q_2 \in \Gamma \pm } \textrm{Res} \ R_{q_1, q_3}({\bf A} (2\Lambda _{q_2}, \delta h, N))_{n \ \textrm{sing.}} e^{-i2\Lambda _{q_3}t} \cr \nonumber \\
&& = \sum_{q_2 \in \Gamma \pm } \frac{1}{(n-1)!} \lim_{p_0 \rightarrow 2\Lambda _{q_2} }\frac{d^{n-1}}{dp_0^{n-1}} \left [(p_0 - 2\Lambda _{q_2} )^n R_{q_1, q_3}({\bf A}(p_0, \delta h , N))_{n \ \textrm{sing.}}e^{-ip_0 t} \right ] \cr\nonumber \\
&& = \sum_{q_3 \in \Gamma \pm } R_{q_1; q_2, q_3}({\bf A}( 2\Lambda _{q_2}, \delta h , N))_{n \ \textrm{amp.}} \frac{1}{(n-1)!} \left [\frac{d^{n-1}e^{-ip_0 t}}{dp_0^{n-1}}\right ] _{p_0 = 2\Lambda _{q_2}} . \label{AppB_Res3}
\end{eqnarray}

Here, $R_{q_1; q_2, q_3}({\bf A}( 2\Lambda _{q_2}, \delta h , N))_{n \ \textrm{amp.}}$ is a part of geometric series containing $n$-th order singular terms with ``amputated'' legs rendering them nonsingular, i.e.,
\begin{equation}
R_{q_1; q_2, q_3}({\bf A}( 2\Lambda _{q_2}, \delta h , N))_{n \ \textrm{amp.}} = \lim_{p_0 \rightarrow 2\Lambda _{q_2} }\left [(p_0 - 2\Lambda _{q_2} )^n R_{q_1, q_3}({\bf A}(p_0, \delta h , N))_{n \ \textrm{sing.}} \right ] . \label{AppB_Res4}
\end{equation}

In the case of a simple pole, we have just demonstrated in (\ref{AppB_Res1}), using a somewhat different procedure, that
\vspace{-8pt}
\begin{eqnarray}
&& \sum_{q_2 \in \Gamma ^\pm } R_{q_1; q_2, q_4}({\bf A}( 2\Lambda _{q_2}, \delta h , N))_{1 \ \textrm{amp.}} = \sum_{q_2 \in \Gamma \pm } \textrm{Res} \ R_{q_1; q_2, q_3}({\bf A} (2\Lambda _{q_2}, \delta h, N))_{\textrm{sim. pole}} \cr\nonumber \\
&& = \frac{2\delta h}{N}\sum_{q_2, q_3 \in \Gamma ^\pm } R_{q_1; q_2, q_3}({\bf \hat {A}}_{q_1}(\delta h , N)) )\cos (\theta_{q_3} + \theta_{q_2}) R_{q_2, q_4 }({\bf \hat {A}}(\delta h , N)) . \label{AppB_Res5}
\end{eqnarray}

In case of higher order singular terms, by repeated use of (\ref{R_A_derivative}) and (\ref{AppB_Res1}), generalization of (\ref{AppB_Res5}), written in matrix form, is straightforward:
\begin{eqnarray}
&& \sum_{q_2 \in \Gamma ^\pm } R_{q_1; q_2, q_3}({\bf A} (2\Lambda _{q_2}, \delta h, N))_{\textrm{n amp.}} \cr \nonumber \\
&& = (-1)^n \sum_{q_2 \in \Gamma ^ \pm } \lim_{p_0 \rightarrow 2\Lambda _{q_2} } \left [(p_0 - 2\Lambda _{q_2} )^{2n}\frac{d^n R_{q_1, q_3}({\bf A}(p_0, \delta h , N))_{\textrm{n \ sing.}} }{(dp_0) ^n} \right ] \cr \nonumber \\
&& = (-1)^n \sum_{q_2 \in \Gamma ^\pm }\lim_{p_0 \rightarrow 2\Lambda _{q_2} } \left \{(p_0 - 2\Lambda _{q_2} )^{2n} \right . \cr \nonumber \\
&& \left. \times \left [{\bf R}({\bf \hat {A}}_{q_1}(p_0, \delta h , N)) \left ( \frac{d {\bf A} (p_0, \delta h, N)) }{dp_0} {\bf R}({\bf \hat {A}}(p_0, \delta h , N) ) \right )^{n} \ \right ]_{q_2, q_3} \right \} . \label{AppB_Res6}
\end{eqnarray}



\begin{thebibliography}{999}
\bibitem{Senegupta1} Sengupta, K.; Powell, S.; Sachdev, S. Quench dynamics across quantum critical points. \emph{Phys. Rev. A} \textbf{2004}, \emph{69}, 053616.
\bibitem{Silva1} Silva, A. The statistics of the work done on a quantum critical system by quenching a control parameter. \emph{Phys. Rev. Lett.} \textbf{2008}, \emph{101}, 120603.
\bibitem{Fagotti1} Fagotti, M.; Calabrese, P. Evolution of entanglement entropy following a quantum quench: Analytic results for the XY chain in a transverse magnetic field. \emph{Phys. Rev. A} \textbf{2008}, \emph{78}, 010306.
\bibitem{Rossini1} Rossini, D.; Silva, A.; Mussardo, G.; Santoro, G. Effective thermal dynamics following a quantum quench in a spin chain. \emph{Phys. Rev. Lett.} \textbf{2009}, \emph{102}, 127204.
\bibitem{Rossini2} Rossini, D.; Suzuki, S.; Mussardo, G.; Santoro, G.E.; Silva, A. Long time dynamics following a quench in an integrable quantum spin chain: Local versus non-local operators and effective thermal behaviour. \emph{Phys. Rev. B} \textbf{2010}, \emph{82}, 144302.
\bibitem{Campos1} Venuti, L.C.; Zanardi, P. Unitary equilibrations: Probability distribution of the Loschmidt echo. \emph{Phys. Rev. A} \textbf{2010}, \emph{81}, 022113.
\bibitem{Gambassi1} Gambassi, A.; Silva, A. Statistics of the Work in Quantum Quenches, Universality and the Critical Casimir Effect. \emph{arXiv} \textbf{2011}
, arXiv:1106.2671v1.
\bibitem{Guo1} Guo, H.L.; Liu, Z.; Fan, H.; Chen, S. Correlation properties of anisotropic XY model with a sudden quench. \emph{Eur. Phys. J. B} \textbf{2011}, \emph{79},~503. 
\bibitem{Canovi1} Canovi, E.; Rossini, D.; Fazio, R.; Santoro, G.E.; Silva, A. Quantum Quenches, Thermalization and Many-Body Localization. \emph{Phys. Rev. B} \textbf{2011}, \emph{83}, 094431.
\bibitem{Campos3} Venuti, L.C.; Jacobson, N.T.; Santra, S.; Zanardi, P. Exact Infinite-Time Statistics of the Loschmidt Echo for a Quantum Quench. \emph{Phys. Rev. Lett.} \textbf{2011}, \emph{107}, 010403.
\bibitem{Calabrese1} Calabrese, P.; Essler, F.H.L.; Fagotti, M. Quantum Quench in the Transverse Field Ising Chain. \emph{Phys. Rev. Lett.} \textbf{2011}, \emph{106}, 227203.
\bibitem{Igloi1} Igl\'oi, F.; Rieger, H. Long-Range correlations in the nonequilibrium quantum relaxation of a spin chain.  \emph{Phys. Rev. Lett.} \textbf{2000}, \emph{85},~3233.
\bibitem{Igloi2} Igl\'oi, F.; Rieger, H. Quantum relaxation after a quench in systems with boundaries. \emph{Phys. Rev. Lett.} \textbf{2011}, \emph{106}, 035701.
\bibitem{Foini1} Foini, L.; Cugliandolo, L.F.; Gambassi, A. Fluctuation-dissipation relations and critical quenches in the transverse field Ising chain. \emph{Phys. Rev. B} \textbf{2011}, \emph{84}, 212404.
\bibitem{Riegler1} Rieger, H.; Igl\'oi, F. Semiclassical theory for quantum quenches in finite transverse Ising chains. \emph{Phys. Rev. B} \textbf{2011}, \emph{84}, 165117.
\bibitem{Polkovnikov1} Polkovnikov, A.; Sengupta, K.; Silva, A.; Vengalattore, M. Colloquium: Nonequilibrium dynamics of closed interacting quantum systems. \emph{Rev. Mod. Phys.} \textbf{2011}, \emph{83}, 863.
\bibitem{Schuricht1} Schuricht, D.; Essler, F.H.L. Dynamics in the Ising field theory after a quantum quench. \emph{J. Stat. Mech.} \textbf{2012}, \emph{2012}, 
P04017.
\bibitem{Calabrese2} Calabrese, P.; Essler, F.H.L.; Fagotti, M. Quantum quench in the transverse field Ising chain: I. Time evolution of order parameter correlators. \emph{J. Stat. Mech.} \textbf{2012}, \emph{2012}, P07016.
\bibitem{Calabrese3} Calabrese, P.; Essler, F.H.L.; Fagotti, M. Quantum quenches in the transverse field Ising chain: II. Stationary state properties. \linebreak \emph{J. Stat. Mech.} \textbf{2012}, \emph{2012}, P07022.
\bibitem{Fagotti2} Fagotti, M. Finite-size corrections versus relaxation after a sudden quench. \emph{Phys. Rev. B} \textbf{2013}, \emph{87}, 165106.
\bibitem{Heyl1} Heyl, M.; Polkovnikov, A.; Kehrein, S. Dynamical Quantum Phase Transitions in the Transverse-Field Ising Model. \emph{Phys. Rev. Lett.} \textbf{2013}, \emph{110}, 135704.
\bibitem{Essler1} Essler, F.H.L.; Fagotti, M. Quench dynamics and relaxation in isolated integrable quantum spin chains. \emph{J. Stat. Mech.} \textbf{2016}, \emph{2016},~064002.
\bibitem{Mitra1} Mitra, A. Correlation functions in the prethermalized regime after a quantum quench of a spin chain. \emph{Phys. Rev. B} \textbf{2013}, \emph{87},~205109.
\bibitem{Marcuzzi1} Marcuzzi, M.; Marino, J.; Gambassi, A.; Silva, A. Prethermalization in a Nonintegrable Quantum Spin Chain after a Quench. \emph{Phys. Rev. Lett.} \textbf{  2013}, \emph{111}, 197203.
\bibitem{Bertini1} Bertini, B.; Fagotti, M. Pre-relaxation in weakly interacting models. \emph{J. Stat. Mech.} \textbf{2015}, \emph{2015}, P07012.
\bibitem{Mitra2} Mitra, A. Quantum Quench Dynamics. \emph{Annu. Rev. Condens. Matter Phys.} \textbf{  2018}, \emph{ 9}, 245. 
\bibitem{Nadkishore1} Nandkishore, R.; Huse, D.A. Many-Body Localization and Thermalization in Quantum Statistical Mechanics. \emph{Annu. Rev. Condens. Matter Phys.} \textbf{2015}, \emph{6}, 15.
\bibitem{Yang1} Yang, Z.-C.; Hamma, A.; Giampaolo, S.M.; Mucciolo, E.R.; Chamon, C. Entanglement complexity in quantum many-body dynamics, thermalization, and localization. \emph{Phys. Rev. B} \textbf{2017}, \emph{96,} 020408.
\bibitem{Zunkovic1} \v{Z}unkovi\v{c}, B.; Silva, A.; Fabrizio, M. Dynamical phase transitions and Loschmidt echo in the infinite-range XY model. \emph{Phil. Trans. R. Soc. A} \textbf{2016}, \emph{374}, 20150160.
\bibitem{Jafari1} Jafari, R. Dynamical Quantum Phase Transition and Quasi Particle Excitation. \emph{Sci. Rep.} \textbf{2019}, \emph{9}, 2871.
\bibitem{Paul1} Paul, S.; Titum, P.; Maghrebi, M. Hidden quantum criticality and entanglement in quench dynamics. \emph{Phys. Rev. Research} \textbf{2024}, \emph{6}, L032003.
\bibitem{Ding1} Ding, C. Dynamical quantum phase transition from a critical quantum quench. \emph{Phys. Rev. B} \textbf{2020}, \emph{102}, 060409.
\bibitem{Porta1} Porta, S.; Cavaliere, F.; Sasseti, M.; Ziani, N.T. Topological classification of dynamical quantum phase transitions in the xy chain. \emph{Sci. Rep.} \textbf{2020}, \emph{10}, 12766.
\bibitem{Lupo1} Lupo, C.; Schir\'o, M. Transient Loschmidt echo in quenched Ising chains. \emph{Phys. Rev. B} \textbf{2016}, \emph{94}, 014310. 
\bibitem{Goussev1} Goussev, A.; Jalabert, R.A.; Pastawski, H.M.; Wisniacki, D.A. 
 Loschmidt echo. \emph{Scholarpedia} \textbf{2012}, \emph{7}, 11687. 
\bibitem{groupLE} Torre, G.; Mari\'{c}, V.; Kui\'{c}, D.; Franchini, F.; Giampaolo, S.M. Odd thermodynamic limit for the Loschmidt echo.  
\emph{Phys. Rev. B} \textbf{2022}, \emph{105}, 184424.
\bibitem{Catalano1} Catalano, A.G.; Giampaolo, S.M.; Morsch, O.; Giovannetti, V.; Franchini, F. Frustrating Quantum Batteries. 
\emph{PRX Quantum} \textbf{2024}, \emph{5},~030319.
\bibitem{Rossini3} Rossini, D.; Calarco, T.; Giovannetti, V.; Montangero, S.; Fazio, R. Decoherence induced by interacting quantum spin baths. \emph{Phys. Rev. A} \textbf{2007}, \emph{75}, 032333.
\bibitem{Goold1} Goold, J.; Fogarty, T.; Gullo, N.L.; Paternostro, M.; Busch, T. Orthogonality catastrophe as a consequence of qubit embedding in an ultracold Fermi gas. \emph{Phys. Rev. A } \textbf{2011}, \emph{84}, 063632.
\bibitem{Knap1} Knap, M.; Shashi, A.; Nishida, Y.; Imambekov, A.; Abanin, D.A.; Demler, E. Time-Dependent Impurity in Ultracold Fermions: Orthogonality Catastrophe and Beyond. \emph{Phys. Rev. X} \textbf{2012}, \emph{2}, 041020.
\bibitem{Knap2} Knap, M.; Kantian, A.; Giamarchi, T.; Bloch, I.; Lukin, M.D.; Demler, E. Probing Real-Space and Time-Resolved Correlation Functions with Many-Body Ramsey Interferometry. \emph{Phys. Rev. Lett.} \textbf{2013}, \emph{111}, 147205.
\bibitem{Dora1} D\'ora, B.; Pollmann, F.; Fort\'agh, J.; Zar\'and, G. Loschmidt Echo and the Many-Body Orthogonality Catastrophe in a Qubit-Coupled Luttinger Liquid. \emph{Phys. Rev. Lett.} \textbf{2013}, \emph{111}, 046402.
\bibitem{Dorner1} Dorner, R.; Clark, S.R.; Heaney, L.; Fazio, R.; Goold, J.; Vedral, V. Extracting Quantum Work Statistics and Fluctuation Theorems by Single-Qubit Interferometry. \emph{Phys. Rev. Lett.} \textbf{2013}, \emph{110}, 230601.
\bibitem{Mazzola1} Mazzola, L.; Chiara, G.D.; Paternostro, M. Measuring the Characteristic Function of the Work Distribution. \emph{Phys. Rev. Lett.} \textbf{2013}, \emph{110}, 230602.
\bibitem{Lieb1} Lieb, E.; Schultz, T.; Mattis, D. Two Soluble Models of an Antiferromagnetic Chain. \emph{Ann. of Phys.} \textbf{1961}, \emph{16}, 407.
\bibitem{Fabio-book} Franchini, F. \emph{An introduction to integrable techniques for one-dimensional quantum systems}; Lecture Notes in Physics; Springer: Cham, Switzerland, 
 2017; Volume 940.
\bibitem{Lancaster} Lancaster, T.; Blundell, S.J. \emph{Quantum Field Theory for the Gifted Amateur}; Oxford University Press: Oxford, UK, 
2014. 
\bibitem{Le Bellac} Le Bellac, M. \emph{Thermal Field Theory}; Cambridge University Press: Cambridge, UK, 1996.
\bibitem{Coleman} Coleman, P. \emph{Introduction to Many-Body Physics}; Cambridge University Press: Cambridge, UK, 2015. 
\bibitem{Dadic1} Dadi\'{c}, I.; Klabu\v{c}ar, D.; Kui\'{c}, D. Direct Photons from Hot Quark Matter in Renormalized Finite-Time-Path QED. \emph{Particles} \textbf{2020}, \emph{3},~676.
\bibitem{Dadic2} Dadi\'{c}, I.; Klabu\v{c}ar, D. Neutrino Oscillations in Finite Time Path Out-of-Equilibrium Thermal Field Theory. \emph{Symmetry } \textbf{2023}, \emph{15},~1970.
\bibitem{Dadic3} Dadi\'{c}, I.; Klabu\v{c}ar, D. Causality and Renormalization in Finite-Time-Path Out-of-Equilibrium $\phi^3$ QFT. \emph{Particles} \textbf{2019}, \emph{2}, 92.
\bibitem{Dadic4} Dadi\'{c}, I. Out of equilibrium thermal field theories: Finite time after switching on the interaction and Wigner transforms of projected functions. \emph{Phys. Rev. D } \textbf{2000}, \emph{63}, 025011.
\bibitem{Damski1} Damski, B.; Rams, M.M. Exact results for fidelity susceptibility of the quantum Ising model: The interplay between parity, system size, and magnetic field. \emph{J. Phys. A: Math. Theor.} \textbf{2014}, \emph{47}, 025303.
\bibitem{Dong1} Dong, J.-J.; Li, P.; Chen, Q.-H. The a-cycle problem for transverse Ising ring. \emph{J. Stat. Mech.} \textbf{2016}, P113102.
\bibitem{Dong2} Dong, J.-J.; Li, P. The a-cycle problem in xy model with ring frustration. \emph{Mod. Phys. Lett. B} \textbf{2017}, \emph{31}, 1750061.
\bibitem{group1} Giampaolo, S.M.; Ramos, F.B.; Franchini, F. The Frustration in being Odd: Area Law Violation in Local Systems. \emph{J. Phys.Commun.} \textbf{2019}, \emph{3}, 081001.
\bibitem{group2} Mari\'{c}, V.; Giampaolo, S.M.; Kui\'{c}, D.;  Franchini, F. The Frustration of being Odd: How Boundary Conditions can destroy Local Order. \emph{New J. Phys.} \textbf{2020}, \emph{22}, 083024,.
\bibitem{group3} Mari\'{c}, V.; Giampaolo, S.M.; Franchini, F. Quantum phase transition induced by topological frustration. \emph{Commun. Phys.} \textbf{2020}, \emph{3},~220.
\bibitem{group4} Mari\'{c}, V.; Franchini, F.; Kui\'{c}, D.; Giampaolo, S.M. Resilience of the topological phases to frustration. \emph{Sci. Rep.} \textbf{2021}, \emph{11}, 6508.
\bibitem{group5} Torre, G.; Mari\'{c}, V.; Franchini, F.; Giampaolo, S.M. Effects of defects in the XY chain with frustrated boundary conditions. \emph{ Phys. Rev. B } \textbf{2021}, \emph{103}, 014429.
\bibitem{group6} Mari\'{c}, V.; Giampaolo, S.M.; Franchini, F. Fate of local order in topologically frustrated spin chains. \emph{ Phys. Rev. B } \textbf{2022}, \emph{105}, 064408.
\bibitem{group7} Mari\'{c}, V.; Torre, G.; Franchini, F.; Giampaolo, S.M. Topological Frustration can modify the nature of a Quantum Phase Transition. \emph{ SciPost Phys.} \textbf{2022}, \emph{12}, 075.
\bibitem{Odavic1} Odavi\'{c}, J.; Haug, T.; Torre, G.; Hamma, A.; Franchini, F.; Giampaolo, S.M. Complexity of frustration: A new source of non-local non-stabilizerness. \emph{SciPost Phys.} \textbf{2023}, \emph{15}, 131.
\bibitem{JWT} P. Jordan, E. Wigner, \"Uber das Paulische \"Aquivalenzverbot. \emph{ Z. Phys.} \textbf{1928}, \emph{47}, 631.
\end{thebibliography}
\end{document}